\documentclass[aip,jmp,reprint,a4paper,amsmath,onecolumn]{revtex4-2}
\usepackage[utf8]{inputenc}
\usepackage{amssymb}
\usepackage{amsthm}
\usepackage{enumitem}
\usepackage{mathrsfs}
\usepackage{multirow} 
\usepackage{stackengine}
\usepackage[colorlinks=true,urlcolor=blue,anchorcolor=blue,citecolor=blue,filecolor=blue,linkcolor=blue,menucolor=blue,linktocpage=true,pdfa=true,unicode]{hyperref}
\usepackage{bm}
\usepackage{lmodern}
\usepackage{graphicx}
\usepackage{ulem}
\newtheorem*{theorem*}{Theorem}

\newcommand{\D}{\mathrm{d}}
\newcommand{\ket}[1]{\left|#1\right\rangle}

\newcommand{\AdS}{\mathrm{AdS}_2}
\newcommand{\SLC}{\widetilde{\mathrm{SL}}(2,\mathbb{R})}

\newcommand{\dom}[1]{\mathrm{Dom}\left(#1\right)}
\newcommand{\domd}[1]{\mathrm{Dom}\left(#1^{\dagger}\right)}

\newcommand{\inpr}[2]{\left\langle #1,#2\right\rangle}
\newcommand{\comment}[1]{}

\renewcommand{\emph}[1]{{\it #1}}

\begin{document}
\title{Dirac field in $\mathrm{AdS}_2$ and representations of $\widetilde{\mathrm{SL}}(2,\mathbb{R})$}

\affiliation{Department of Mathematics, University of York, Heslington, York, YO10 5DD, United Kingdom}
\author{David Serrano Blanco}
\email{dsb523@york.ac.uk}
\affiliation{Department of Mathematics, University of York, Heslington, York, YO10 5DD, United Kingdom}

\date{\today}

\begin{abstract}
We study the solutions to the Dirac equation for the massive spinor field in the universal covering space of two--dimensional anti--de Sitter space. For certain values of the mass parameter, we impose a suitable set of boundary conditions which make the spatial component of the Dirac operator self--adjoint. Then, we use the  transformation properties of the spinor field under the isometry group of the theory, namely, the universal covering group of $\mathrm{SL}(2,\mathbb{R})$, in order to determine which self--adjoint boundary conditions are invariant under this group. We identify the corresponding  solution spaces with unitary irreducible representations of this group using the classification given by Pukanzki,~\cite{puk} and determine which of these correspond to invariant positive-- and negative--frequency subspaces and, hence, in a vacuum state invariant under the isometry group. Finally, we examine the cases where the self--adjoint boundary condition leads to an invariant theory with non--invariant vacuum state and determine the unitary representation to which the vacuum state belongs.
\end{abstract}

\maketitle

\section{Introduction}\label{Intro}

Field theory in the covering space of the anti--de Sitter spacetime has played a central role in  theoretical and mathematical physics in the last two decades, mainly due to its relevance in the $\mathrm{AdS}/\mathrm{CFT}$ correspondence.\cite{maldacena} Interest in classical and quantum theories on this spacetime has gone well beyond its initial connection to this framework from the viewpoint of string theory and has resulted in the investigation of their properties in a wide variety of different contexts.\cite{dappiaggi,morley} 

The $N$--dimensional anti--de Sitter manifold is not globally hyperbolic, therefore, global solutions to hyperbolic equations describing the field dynamics cannot be found only by specifying initial data. Nevertheless, by providing additional asymptotic conditions to the solutions at the spatial boundary it is possible to define sensible and deterministic field dynamics. Several approaches can be taken to justify particular choices of boundary conditions for field theories in $\mathrm{AdS}$ spacetimes depending on the context in which these theories are analyzed.~\cite{avis,breit,sak} One in particular, taken by Ishibashi and Wald~\cite{ishi1,ishi2,ishi3} for integer spin fields (scalar, vector and symmetric tensor fields) defined on the universal cover of $N$--dimensional anti--de Sitter space with $N\geq 3$, focuses on finding a family of boundary conditions by applying the theory of self--adjoint extensions to the radial component of the spatial operator that all these fields share in common. The result is a family of theories parametrized by a real number, and the familiar boundary conditions like Dirichlet, Neumann and Robin boundary conditions emerge as special cases of certain values of this parameter. In a previous work,~\cite{HigSerr2} the author, in collaboration, applied this approach to the particular case of a minimally coupled, non--interacting scalar field obeying the Klein--Gordon equation in the universal covering space of two--dimensional anti--de Sitter space. (Here and hereafter we will refer to the universal cover of the $N$--dimensional anti-de Sitter manifold as anti--de Sitter space, $\mathrm{AdS}_{N}$.) In this spacetime the spatial coordinate is no longer radial and thus, unlike in the higher-dimensional case studied by Ishibashi and Wald, the spatial boundary consists of two disjoint points. Due to this fact the self-adjoint extensions for the spatial Klein--Gordon operator in $\mathrm{AdS}_2$ turned out to be richer than in the higher--dimensional case, parametrized by a $2\times 2$ unitary matrix instead of a real number. We also noted that not all of the theories that were obtained from this approach may be of physical interest. Depending on the context in which such theories are analyzed, different arguments may be given for choosing a particular theory over the others. The criteria we chose was to require the invariance of the resulting theories under the isometry group of the spacetime. Among the family of different consistent theories arising from self--adjoint boundary conditions, we found those whose positive--frequency solutions formed a unitary irreducible representation of the symmetry group of $\mathrm{AdS}_2$, \textit{i.e.}\ the universal covering group of $\mathrm{SL}(2,\mathbb{R})$, denoted by $\widetilde{\mathrm{SL}}(2,\mathbb{R})$. The classification of all unitary irreducible representations of this group up to isomorphisms has been obtained by Pukanzki~\cite{puk} using an analysis similar to that of Bargmann~\cite{barg} for the case of $\mathrm{SL}(2,\mathbb{R})$ and is well known. \cite{lang,knapp,har,rep,kitaev} Each representation is characterized by two parameters which can be associated to the mass and energy spectrum of the Klein-Gordon equation in $\AdS$. The mode solutions obtained by requiring invariance under the action of $\SLC$ agreed with the results of Sakai and Tanii.~\cite{sak} In their work they studied free scalar and spinor field theories in $\AdS$ with the boundary conditions for the mode solutions being determined by imposing the vanishing of energy flux at the conformal boundary. Furthermore, the theory of self--adjoint extensions has been applied to a Dirac field in $\mathrm{AdS}_{4}$ in a slightly different way by Bachelot,~\cite{bach} where certain boundary conditions are pointed out.

In this paper a spinor field of mass $M$ obeying the Dirac equation in $\mathrm{AdS}_2$ is analyzed. Following the approach taken for the scalar field case, we will first obtain the self-adjoint extensions of the spatial component of the Dirac operator, and express them in terms of boundary conditions at the spatial boundary for the mode solutions. The type of self-adjoint boundary conditions depends on the absolute value of the mass of the field. If $|M|$ is sufficiently large, the boundary conditions are uniquely determined by requiring the solutions to the Dirac equation to be normalizable with respect to the Dirac inner product, which is equivalent to the fact that the spatial Dirac operator for this mass range is essentially self--adjoint and thus, has a unique self--adjoint extension. On the other hand, in a certain range of low mass parameter, similarly to the case of a scalar field, the self--adjoint extensions will be parametrized by a $2\times 2$ unitary matrix. We will then determine which of those boundary conditions result in invariant mode solutions under the infinitesimal action of $\SLC$, which can be realized through a certain Lie derivative operator defined on spinor solutions. We will then find which of the resulting invariant mode solution spaces admit an invariant positive-- or negative--frequency subspace and hence, result in a vacuum state invariant under the $\SLC$ action. Finally, we analyze the cases for which the invariant boundary conditions result in modes which do not admit this frequency spectrum splitting and thus, describe quantum theories with non-invariant vacuum sectors.

The rest of the paper is organized as follows. In Sec.~\ref{AdSsection} we describe the geometry of $\mathrm{AdS}_2$ together with the spin structure needed to describe spinors on this manifold, and briefly summarize the classification of unitary irreducible representations of $\SLC$ following the conventions of Kitaev.~\cite{kitaev} In Sec.~\ref{DiracEqSec} we begin the analysis of the solutions to the Dirac equation, presenting the general solutions of the associated spatial equation for different ranges of the mass parameter. In Sec.~\ref{SelfAdjSec} we apply the theory of self--adjoint extensions due to Weyl~\cite{weyl} and von~Neumann~\cite{neu} to the associated spatial differential operator in order to obtain the self--adjoint boundary conditions. In Sec.~\ref{InvariantBC-Sec} we determine which of these boundary conditions respect the $\widetilde{\mathrm{SL}}(2,\mathbb{R})$ symmetry, and in Sec.~\ref{InvSec} we present the explicit mode solutions for each of these boundary conditions. In Sec.~\ref{UIR-Sec} we determine which of the sets of mode solutions characterized by the invariant self--adjoint boundary conditions form a unitary irreducible representation of $\SLC$, and determine which among these result in invariant positive-- and negative--frequency subspaces. In Sec.~\ref{NonInv-Sec} we analyze the cases where the boundary condition is invariant but the vacuum state is not, and determine the unitary irreducible representation this state belongs to. We summarize our results in Sec.~\ref{Concl}.
In Appendix~\ref{App2} we review a free quantum Dirac field with a stationary vacuum state in a general static spacetime. In Appendix~\ref{App1} we present some technical details regarding the asymptotic behavior of the spatial solutions for certain values of the mass parameter.

\section{Geometry of anti-de Sitter space and spin structure}\label{AdSsection}

Two--dimensional anti--de Sitter space is the two--dimensional hyperboloid embedded in three--dimensional flat space with Lorentzian signature $(-,-,+)$ and coordinates $X^0,X^1,X^2$ given by the equation
\begin{align}\label{hyperboloid}
\left(X^0\right)^2+\left(X^1\right)^2-\left(X^2\right)^2=1\,.
\end{align}

By choosing the coordinate system 
\begin{subequations}\label{coords}
\begin{align}
X^0&=\sec\rho\cos\,t\,,\\
X^1&=\sec\rho\sin\,t\,,\\
X^2&=\tan\rho\,,
\end{align}
\end{subequations}
with $\rho\in(-\pi/2,\pi/2)$, and $t\in[-\pi,\pi)$, the induced metric for anti--de Sitter space with coordinates  $\left(x^{\mu}\right)=\left(x^{0},x^{1}\right)=(t,\rho)$, takes the form
\begin{align}\label{metric}
\D s^2=g_{\mu\nu}\D x^{\mu}\D x^{\nu}=\sec^2\rho\left(-\D t^2+\D\rho^2\right)\,,
\end{align}
which is conformally flat. We see that this metric is of the form~\eqref{StaticMetric} with $D=2$ and $N=\sec\rho=(g_{ij})$. The universal covering space of the two--dimensional anti--de Sitter space, $\AdS$, is the pseudo--Riemannian manifold with metric given by Eq.~\eqref{metric} with the coordinate $t$ taking all real values. 

We now introduce the local orthonormal frame $\left\{e_{a}\right\}$, where $a=0,1$, denotes the local Lorentz index, whose non--zero spacetime components $e_{a}^{~\mu}$ are given by
\begin{align}\label{zweibein}
e_{0}^{~0}=\cos\rho=e_{1}^{~1}\,.
\end{align}
The frame field components satisfy the following  orthonormality relations:
\begin{align}\label{zweibein}
g_{\mu\nu}e_{a}^{~\mu}e_{b}^{~\nu}=\eta_{ab}\,,\hspace{.5cm}e_{a}^{~\mu}e_{~\mu}^{b}=\delta_a^b\,,\hspace{.5cm}e_{a}^{~\mu}e_{~\nu}^{a}=\delta_\nu^\mu\,,
\end{align}
where $\eta=\mathrm{diag}(-1,1)$ is the flat metric, and the functions $e_{~\mu}^{a}$ denote the spacetime components of the co--frame fields $\left\{\hat{e}^{a}\right\}$. With respect to this local frame we take the associated connection $1$--form $\omega^{a}_{~b}$ to be defined by
\begin{align}\label{spinconnection}
\omega^{a}_{~b\,\mu}=\left(\partial_{\mu}e^{~\lambda}_{b}+\Gamma^{\lambda}_{~\mu\nu}e_{b}^{~\nu}\right)e^{a}_{~\lambda}\,,
\end{align}
where $\Gamma^{\lambda}_{~\mu\nu}$ are the components of the metric connection for this coordinate system. Using Eq.~\eqref{metric} it follows that the only non--zero components of the metric connection are $\Gamma^{0}_{~01}=\Gamma^{1}_{~00}=\Gamma^{1}_{~11}=\tan\rho$, and the non--zero components of the connection $1$--form are $\omega^{0}_{~1\,0}=\omega^{1}_{~0\,0}=\tan\rho$. Following a convention similar to that of Sakai and Tanii,~\cite{sak} we will use the $2$--dimensional representation of the  $2\times 2$--gamma matrices $\gamma^{a}$ given by
\begin{align}\label{GammaMatrices}
\gamma^{0}=\begin{pmatrix}
0 & i\\
i & 0
\end{pmatrix}\,,\hspace{.5cm}\gamma^{1}=\begin{pmatrix}
-1 & 0 \\
0 & 1
\end{pmatrix}\,,
\end{align}
which satisfy the anticommutation relation $\{\gamma^{a},\gamma^{b}\}=2\eta^{ab}\mathbb{I}$, with $\mathbb{I}$ denoting the identity matrix. Furthermore, we have $(\gamma^{0})^{\dagger}=-\gamma^{0}$ and $(\gamma^1)^{\dagger}=\gamma^1$. We also define the matrices  $\Sigma^{ab}=[\gamma^{a},\gamma^{b}]/4$,  for which only 
\begin{align}\label{sigmamatrix}
\Sigma^{01}=\frac{1}{2}\begin{pmatrix}
0 & i\\
-i & 0
\end{pmatrix}\,,
\end{align}
and $\Sigma^{10}=-\Sigma^{01}$ are non--trivial in the two--dimensional case we are considering. In this representation the charge conjugation matrix $C$ is given by $C=2z\Sigma^{01}$ for any $z\in\mathbb{C}$ on the unit circle. We choose $z=-1$.

Spinor fields will be regarded as elements of the space $C^{\infty}(\AdS,\mathbb{C}^{2})$, of $\mathbb{C}^{2}$--valued smooth functions on anti--de Sitter space. Using the convention consistent with the definition in Eq.~\eqref{spinconnection}, the spinor covariant derivative, $\nabla_{\mu}$, is given by
\begin{align}\label{covariantder}
\nabla_{\mu}=\partial_{\mu}+\frac{1}{2}\omega_{ab\,\mu}\Sigma^{ab}\,,
\end{align}
where $\omega_{ab\,\mu}=\eta_{ac}\omega^{c}_{~b\,\mu}$. The Dirac operator on this manifold is then defined as $D:=\gamma^{\mu}\nabla_{\mu}=e^{~\mu}_{a}\gamma^{a}\nabla_{\mu}$ and, hence, the Dirac equation for a spinor field $\psi\in C^{\infty}(\AdS,\mathbb{C}^2)$ with mass $M\in\mathbb{R}$ is written as
\begin{align}\label{DiracEq1}
\gamma^{\mu}\nabla_{\mu}\psi-M\psi=0\,.
\end{align}
Considering the Dirac adjoint $\psi^{*}=\psi^{\dagger}\gamma^{0}$,  with $\psi^{\dagger}$ denoting the conjugate transpose, it follows that $\nabla_{\mu}\left(\psi^{*}\gamma^{\mu}\psi\right)=0$ when $\psi$ is a solution to the Dirac equation, hence, the quantity
\begin{align}\label{current}
\int_{-\frac{\pi}{2}}^{\frac{\pi}{2}}\psi^{\dagger}\psi\sqrt{-g}\,e_{0}^{~0}\,\D\rho=\int_{-\frac{\pi}{2}}^{\frac{\pi}{2}}\psi^{\dagger}\psi\frac{\D\rho}{\cos\rho}\,,
\end{align}
where $g$ stands for the determinant of the metric in Eq.~\eqref{metric}, is conserved. Therefore, given two solutions $\psi_1,\psi_2\in C^{\infty}(\AdS,\mathbb{C}^2)$ of Eq.~\eqref{DiracEq1}, their inner product, given by
\begin{align}\label{innerp1}
\inpr{\psi_1}{\psi_2}_{D}=\int_{-\frac{\pi}{2}}^{\frac{\pi}{2}}\psi_{1}(t,\rho)^{\dagger}\psi_{2}(t,\rho) \frac{\D\rho}{\cos\rho}\,,
\end{align}
is time--independent. 

The isometry group of $\AdS$ is $\SLC$, that is, the universal covering group of $\mathrm{SL}(2,\mathbb{R})$. In later sections we will use the invariance of the solutions to Eq.~\eqref{DiracEq1} under the infinitesimal action of this group as a criterion for preferred theories, therefore we will obtain the associated infinitesimal operators acting on the space of solutions. The Killing vector fields leaving the metric in Eq.~\eqref{metric} invariant are given by
\begin{subequations}\label{Killingvectors}
\begin{align}
\xi_{0}&=\partial_{t}\,,\label{Kill0}\\
\xi_{1}&=\cos\,t\sin\rho\,\partial_{t}+\sin\,t\cos\rho\,\partial_{\rho}\,,\label{Kill1}\\
\xi_{2}&=-\sin\,t\sin\rho\,\partial_{t}+\cos\,t\cos\rho\,\partial_{\rho}\,,\label{Kill2}
\end{align}
\end{subequations}
which satisfy the $\mathfrak{sl}(2,\mathbb{R})$ commutation relations, namely,
\begin{align}\label{Commute-Kill}
\left[\xi_{0},\xi_{1}\right]=\xi_{2}\,,\hspace{.5cm}\left[\xi_{0},\xi_{2}\right]=-\xi_{1}\,,\hspace{.5cm}\left[\xi_{1},\xi_{2}\right]=-\xi_{0}\,.
\end{align}
For a spinor field $\psi\in C^{\infty}(\AdS,\mathbb{C}^{2})$, the spinorial Lie derivative~\cite{collas,vas} in the direction of an arbitrary vector field $\xi$, is given by the expression
\begin{align}\label{LieDer}
\mathcal{L}_{\xi}\psi=\xi^{\mu}\nabla_{\mu}\psi+\frac{1}{4}(\nabla_{\mu}\xi_{\nu})\gamma^{\mu}\gamma^{\nu}\psi\,,
\end{align}
so the spinorial Lie derivatives in the direction of the Killing vector fields in Eq.~\eqref{Killingvectors} define an infinitesimal action of $\SLC$ on the space of spinor fields, and explicitly read
\begin{subequations}\label{SpinLieDer}
\begin{align}
\mathcal{L}_{\xi_0}&=\partial_{t}\,,\label{SpinLieDer0}\\
\mathcal{L}_{\xi_{1}}&=\cos t\,\sin\rho\,\partial_{t}+\sin t\,\cos\rho\,\partial_{\rho}+\cos t\,\cos\rho\,\Sigma^{01}\,,\label{SpinLieDer1}\\
\mathcal{L}_{\xi_{2}}&=-\sin t\,\sin\rho\,\partial_{t}+\cos t\,\cos\rho\,\partial_{\rho}-\sin t\,\cos\rho\,\Sigma^{01}\,.\label{SpinLieDer2}
\end{align}
\end{subequations}
From these operators we can construct the associated time--translation operator $\mathcal{L}_{0}:=i\mathcal{L}_{\xi_{0}}$ and the ladder operators
\begin{align}\label{ladderop1}
\mathcal{L}_{\pm}&:=\mathcal{L}_{\xi_{1}}\pm i\mathcal{L}_{\xi_{2}}\,,\nonumber\\
&=e^{\mp it}\left(\pm i\cos\rho\,\partial_{\rho}+\sin\rho\,\partial_{t}+\cos\rho\,\Sigma^{01}\right)\,,
\end{align}
which will be more convenient to use once we obtain a mode decomposition of the solutions. We note that the commutation relations between these operators are given by
\begin{align}\label{Comm-LadderOps}
\left[\mathcal{L}_0,\mathcal{L}_{\pm}\right]=\pm\mathcal{L}_{\pm}\,,\hspace{.5cm}\left[\mathcal{L}_{+},\mathcal{L}_{-}\right]=2\mathcal{L}_{0}\,.
\end{align}

The solution space of the Dirac equation with mass $M$ can be identified with the unitary irreducible representations (UIR's) of the group $\SLC$. Here we give a brief description of the classification of all UIR's up to isomorphism due to Pukanzki~\cite{puk} adopting a notation and conventions similar to those in the notes by Kitaev.~\cite{kitaev} We describe the classification directly in terms of the action of $\SLC$ on the space of spinors $C^{\infty}(\AdS,\mathbb{C}^2)$. A more detailed discussion on how these representations are identified in a more general Hilbert space can be found in our previous work~\cite{HigSerr2} and standard  references.~\cite{har,rep}
On the space of spinors, $C^{\infty}(\AdS,\mathbb{C}^2)$, the quadratic Casimir operator is given in terms of the operators in Eq.~\eqref{Comm-LadderOps} by
\begin{align}
Q&=\mathcal{L}_{0}^2+\frac{1}{2}\left(\mathcal{L}_{+}\mathcal{L}_{-}+\mathcal{L}_{-}\mathcal{L}_{+}\right)\,,
\end{align}
and thus, if the representation is irreducible, by Schur's lemma, this operator acts as multiplication by a number $q$. Furthermore, if the representation is unitary, then the ladder operators $\mathcal{L}_{\pm}$ in Eq.~\eqref{ladderop1} satisfy $\mathcal{L}_{\pm}^{\dagger}=-\mathcal{L}_{\mp}$ and the Casimir eigenvalue, $q$, is a real number. A direct calculation shows that the squared Dirac operator $D^2=\gamma^\mu\nabla_\mu\gamma^\nu\nabla_\nu$ is related to the Casimir element by $D^2=Q+(1/4)\mathbb{I}$ and thus, the mass $M$ of the spinor solutions is related to the Casimir eigenvalue by $M^2=q+1/4$. With this parametrization, the UIR's of $\SLC$ are labeled by the pair $(M,\mu)$, with $M\in\mathbb{R}$ or $M=is$, with $0<s\in\mathbb{R}$, and where $\mu\in\mathbb{R}/\mathbb{Z}$ is defined through the eigenvalue, $e^{-2\pi i\mu}$, of the central element $\exp(2\pi\mathcal{L}_{\xi_{0}})$. Each representation is realized as the Hilbert space spanned by the simultaneous eigenvectors of $Q$ and the operator $\mathcal{L}_{0}$, the latter of which has eigenvalue $\omega\in\mu+\mathbb{Z}$. The non--trivial UIR's of $\SLC$ are:
\begin{enumerate}
\item {\bf Discrete series representations:} $\mathscr{D}_{1/2+M}^{\pm}$ for $M>0$, with $\mu=\pm(1/2+M)$, and $\omega=\pm(1/2+M+n)$, respectively, where $n\in\mathbb{N}\cup\{0\}$; $\mathscr{D}_{1/2-M}^{\pm}$ for $0<M<1/2$, with $\mu=\pm(1/2-M)$ and $\omega=\pm(1/2-M+n)$, respectively, where $n\in\mathbb{N}\cup\{0\}$. The Casimir eigenvalue satisfies $q>-1/4$.
\item {\bf Principal series representations:} $\mathscr{P}_{is}^{\mu}$ for $M=is$, with $s\in\mathbb{R}^+$ and $- 1/2<\mu \leq 1/2$, or $s=0$ and $|\mu|<1/2$. For both we have $\omega=\mu+k$, where $k\in\mathbb{Z}$. The Casimir eigenvalue satisfies $q\leq-1/4$.
\item {\bf Complementary series representations:} $\mathscr{C}_{1/2+M}^{\mu}$ for $-1/2<M<1/2$, with  $|\mu|<1/2+M$, and $\omega=\mu+k$, where $k\in\mathbb{Z}$. The Casimir eigenvalue satisfies $-1/4<q<0$.
\item {\bf Mock-discrete series representations:} $\mathscr{D}_{1/2}^{\pm}$ for $M=0$ with $\mu=\pm 1/2$, and $\omega=\pm(1/2+n)$, respectively, with $k\in\mathbb{N}\cup\{0\}.$ The Casimir eigenvalue is $q=-1/4$.
\end{enumerate}

\section{Dirac field in $\mathrm{AdS}_2$}\label{DiracEqSec}

We now proceed to find normalizable solutions of Eq.~\eqref{DiracEq1} with respect to the inner product given by Eq.~\eqref{innerp1}. Using the covariant derivative defined in Eq.~\eqref{covariantder}, we write Dirac's equation in global coordinates $(t,\rho)$ as
\begin{align}\label{DiracEq2}
\cos\rho\left[\gamma^{0}\partial_{t}+\gamma^{1}\left(\partial_{\rho}+\frac{1}{2}\tan\rho\right)\right]\psi(t,\rho)=M\psi(t,\rho)\,.
\end{align}
It will be convenient to define the two--component spinor $\Psi$ via the relation
\begin{align}\label{prefactor1}
\psi(t,\rho)=(\cos\rho)^{\frac{1}{2}}\Psi(t,\rho)\,,
\end{align}
so that Eq.~\eqref{DiracEq2} is equivalent to the equation for the spinor $\Psi$ given by
\begin{align}\label{DiracEq3}
\left(\gamma^{0}\partial_t+\gamma^{1}\partial_\rho\right)\Psi(t,\rho)=M\sec\rho\,\Psi(t,\rho)\,.
\end{align}
A simple substitution shows that if $\Psi$ is a solution of this equation with mass $M$, then $i\Sigma^{01}\Psi$ is a solution with mass $-M$. Then, without loss of generality, we will only consider solutions to Eq.~\eqref{DiracEq3} with $M\geq 0$.

We are interested in the description of the solutions of this equation in terms of positive-- and negative--frequency mode spinors with respect to the timelike Killing vector field $\xi_{0}$ in Eq.~\eqref{Kill0}. We will then consider solutions of the form
\begin{align}\label{modedecomp1}
\Psi(t,\rho)=\Phi_{\omega}(\rho)e^{-i\omega t}\,, \hspace{.5cm} \omega>0\,,
\end{align}
where $\Phi_{\omega}\in C^{\infty}((-\pi/2,\pi/2),\mathbb{C}^{2})$ will be referred to as the \emph{spatial} component of the spinor $\Psi$. From Eq. \eqref{DiracEq3}, it follows that this spatial component satisfies 
\begin{align}\label{SpatialDirac1}
\mathbb{D}\,\Phi_{\omega}(\rho)=\omega\Phi_{\omega}(\rho)\,,
\end{align}
where we have defined the operator
\begin{align}\label{OpD}
\mathbb{D}:=\begin{pmatrix}
0 & -\frac{\D}{\D\rho}+M\sec\rho\\[.5em]
\frac{\D}{\D\rho}+M\sec\rho & 0
\end{pmatrix}\,.
\end{align}
We will concern ourselves with the precise description of the domain of this operator in short. We now project Eq.~\eqref{DiracEq3} onto the components of the spinor $\Phi_{\omega}$ on this representation. Define
\begin{align}\label{components1}
\Phi_{\omega}=\begin{pmatrix}
\Phi^{(1)}_{\omega}\\
\Phi^{(2)}_{\omega}
\end{pmatrix}\,,
\end{align}
with $\Phi_{\omega}^{(1)},\Phi_{\omega}^{(2)}$ complex--valued functions on the interval $(-\pi/2,\pi/2)$. This results in the coupled system of equations given by
\begin{subequations}\label{CoupDir1}
\begin{align}
\frac{\D}{\D\rho}\Phi_{\omega}^{(1)}(\rho)+M\sec\rho\, \Phi_{\omega}^{(1)}(\rho)&=\omega \Phi_{\omega}^{(2)}(\rho)\,,\label{ComponentEq1}\\
-\frac{\D}{\D\rho}\Phi_{\omega}^{(2)}(\rho)+M\sec\rho\, \Phi_{\omega}^{(2)}(\rho)&=\omega \Phi_{\omega}^{(1)}(\rho)\,.\label{ComponentEq2}
\end{align}
\end{subequations}
We note that if $\Phi_\omega$ is a solution of Eq.~\eqref{SpatialDirac1} with $\omega>0$, then the charge conjugate spinor $\Phi_{\omega}^{c}:=C(\gamma^{0})^{T}\overline{\Phi_{\omega}}$, where $C=-2\Sigma^{01}$ is the charge conjugation matrix, is a solution of the same equation with $-\omega$.

Now, eliminating $\Phi_{\omega}^{(2)}$ in Eq.~\eqref{CoupDir1} gives the second order equation
\begin{align}\label{SecondOrdDir}
\frac{\D^2}{\D\rho^2}\Phi_{\omega}^{(1)}(\rho)+\left[\omega^2+M\sec\rho\,\tan\rho-M^2\sec^2\rho\right]\Phi_{\omega}^{(1)}(\rho)=0\,.
\end{align}
A general solution to this equation when $M-1/2\notin\mathbb{N}\cup\{0\}$ is given in terms of the Gaussian hypergeometric functions~\cite{nist} and reads
\begin{align}\label{SolComp1}
\Phi_{\omega}^{(1)}(\rho)=&(2M+1)C_1\,\sigma(\rho)^{M}F\left(\omega,-\omega;\frac{1}{2}+M;\frac{1-\sin\rho}{2}\right)\nonumber\\
&+\omega\,C_2\,\cos\rho\,\sigma(\rho)^{-M}F\left(1+\omega,1-\omega;\frac{3}{2}-M;\frac{1-\sin\rho}{2}\right)\,,
\end{align}
where we have defined
\begin{align}\label{sol-pref}
\sigma(\rho):=\left(\frac{1-\sin\rho}{1+\sin\rho}\right)^{\frac{1}{2}}\,,
\end{align}
and with $C_1,C_2\in\mathbb{C}$ constants. Using this solution we now define the second spinor component $\Phi_{\omega}^{(2)}$ through Eq.~\eqref{ComponentEq1}. Applying standard recursion relations for the hypergeometric functions it can readily be verified that the second component is given by
\begin{align}\label{SolComp2}
\Phi_{\omega}^{(2)}(\rho)=&\omega\,C_1\cos\rho\,\sigma(\rho)^{M}F\left(1+\omega,1-\omega;\frac{3}{2}+M;\frac{1-\sin\rho}{2}\right)\nonumber\\
&+(2M-1)C_2\,\sigma(\rho)^{-M}F\left(\omega,-\omega;\frac{1}{2}-M;\frac{1-\sin\rho}{2}\right)\,.
\end{align}
If $M=1/2+k$, with $k\in\mathbb{N}\cup\{0\}$, then it can be shown that the general solutions are instead given by
\begin{subequations}\label{SolCompklog}
\begin{align}
\Phi_{\omega}^{(1)}(\rho)=&\sigma(\rho)^{\frac{1}{2}}\left[C_1\left(\mathsf{P}_{\omega}^{-k}(\sin\rho)+\mathsf{P}_{\omega-1}^{-k}(\sin\rho)\right)+C_2\left(\mathsf{Q}_{\omega}^{-k}(\sin\rho)+\mathsf{Q}_{\omega-1}^{-k}(\sin\rho)\right)\right]\,,
\label{SolCompklog1}\\
\Phi_{\omega}^{(2)}(\rho)=&\sigma(\rho)^{-\frac{1}{2}}\left[C_1\left(\mathsf{P}_{\omega-1}^{-k}(\sin\rho)-\mathsf{P}_{\omega}^{-k}(\sin\rho)\right)+C_2\left(\mathsf{Q}_{\omega-1}^{-k}(\sin\rho)-\mathsf{Q}_{\omega}^{-k}(\sin\rho)\right)\right]\,,\label{SolCompklog2}
\end{align}
\end{subequations}
where $\mathsf{P}_{\nu}^{\mu}$ and $\mathsf{Q}_{\nu}^{\mu}$ are Ferrers functions (analytic continuations of associated Legendre functions) of the first and second kind,~\cite{nist} respectively. These are given by
\begin{subequations}\label{Ferrers}
\begin{align}
\mathsf{P}_{\nu}^{\mu}(x)&=\frac{1}{\Gamma(1-\mu)}\left(\frac{1+x}{1-x}\right)^{\mu/2}F\left(\nu+1,-\nu;1-\mu;\frac{1-x}{2}\right)\,,\label{Gen-FerrersP}\\
\mathsf{Q}_{\nu}^{\mu}(x)&=\frac{\pi}{\sin\mu\pi}\left(\cos\mu\pi\,\mathsf{P}_{\nu}^{\mu}(x)-\frac{\Gamma(\nu+\mu+1)}{\Gamma(\nu-\mu+1)}\mathsf{P}_{\nu}^{-\mu}(x)\right)\,.\label{Gen-FerrersQ}
\end{align}
\end{subequations}
The functions $\mathsf{Q}_{\nu}^{-k}$ with $k\in\mathbb{N}\cup\{0\}$ can be defined by substituting Eq.~14.9.3 in the DLMF~\cite{nist} into Eq.~\eqref{Gen-FerrersQ} and taking the limit $\mu\to-k$. Solutions for $M=1/2$ reduce to Legendre functions by means of the relation $\mathsf{P}_{\omega}^{0}(x)=\mathsf{P}_{\omega}(x)$. We also note that the solutions for the massless spinor field,  can be directly obtained from Eqs.~\eqref{CoupDir1} and~\eqref{SecondOrdDir} by setting  $M=0$, in which case the components are simply given by
\begin{subequations}\label{MasslessComp1}
\begin{align}
\Phi_{\omega,M=0}^{(1)}(\rho)&=\tilde{C}_1\cos\omega\rho+\tilde{C}_2\sin\omega\rho\,,\label{MasslessSol1}\\
\Phi_{\omega,M=0}^{(2)}(\rho)&=-\tilde{C}_1\sin\omega\rho+\tilde{C}_2\cos\omega\rho\,,\label{MasslessSol2}
\end{align}
\end{subequations}
for some $\tilde{C_1},\tilde{C_2}\in\mathbb{C}$.

The inner product for the full spinor solutions $\psi$ in Eq.~\eqref{innerp1} induces a time--independent inner product for the re--scaled spinors $\Psi$ defined though Eq.~\eqref{prefactor1}. This is given by
\begin{align}\label{innerp3}
\inpr{\Psi_1}{\Psi_2}_{2}:=\inpr{(\cos\rho)^{1/2}\cdot\Psi_{1}}{(\cos\rho)^{1/2}\cdot\Psi_{2}}_{D}\,,
\end{align}
for any two solutions $\Psi_1,\Psi_2$ of Eq.~\eqref{DiracEq2}. The norm squared of the spinor $\Psi$ is given by $\left|\!\left|\Psi\right|\!\right|^{2}_{2}=\inpr{\Psi}{\Psi}_{2}$. 
Assuming that the solutions $\Psi_1$ and $\Psi_2$ are of the form given by Eq.~\eqref{modedecomp1}, we can write $\Psi_{1}(t,\rho)=\Phi_{\omega_1}(\rho)e^{-i\omega_1 t}$ and $\Psi_{2}(t,\rho)=\Phi_{\omega_2}(\rho)e^{-i\omega_2 t}$ for some $\omega_1,\omega_2>0$. Then, Eq.~\eqref{innerp1} and the time--independence of Eq.~\eqref{innerp3} imply that $\inpr{\Psi_1}{\Psi_2}_{2}=\inpr{\Phi_{\omega_1}}{\Phi_{\omega_1}}$, where we have defined 
\begin{align}\label{innerp2}
\inpr{\Phi_{\omega_1}}{\Phi_{\omega_2}}&:=\int_{-\frac{\pi}{2}}^{\frac{\pi}{2}}\Phi_{\omega_{1}}^{\dagger}(\rho)\Phi_{\omega_2}(\rho)\D\rho\,,\nonumber\\
&=\int_{-\frac{\pi}{2}}^{\frac{\pi}{2}}\left(\overline{\Phi_{\omega_1}^{(1)}(\rho)}\Phi_{\omega_2}^{(1)}(\rho)+\overline{\Phi_{\omega_1}^{(2)}(\rho)}\Phi_{\omega_2}^{(2)}(\rho)\right)\D\rho\,,
\end{align}
with $\overline{\Phi(\rho)}$ denoting complex conjugation. Hence, this equation defines an inner product for solutions of Eq.~\eqref{DiracEq3}. The associated norm $\left|\!\left|\Phi_{\omega}\right|\!\right|^{2}_{2}=\inpr{\Phi_\omega}{\Phi_\omega}$ reduces to the sum of the $L^{2}$--norms of the component functions $\Phi_{\omega}^{(1)}$ and $\Phi_{\omega}^{(2)}$. We can then consider the Hilbert space of square--integrable spatial spinors, $L^{2}([-\pi/2,\pi/2],\mathbb{C}^2)$, as the completion of $C^{\infty}((-\pi/2,\pi/2),\mathbb{C}^2)$ with respect to the norm induced by Eq.~\eqref{innerp2}. Therefore, the normalization of the solutions to the spatial component of the Dirac equation is equivalent to the $L^{2}$--normalization of the component functions appearing in Eqs.~\eqref{SolComp1},~\eqref{SolComp2},~\eqref{SolCompklog} and~\eqref{MasslessComp1}.

Considering these facts, we are led to study the properties of the operator $\mathbb{D}$ to determine which solutions of Eq.\eqref{DiracEq3} result in a well--defined initial value problem for Eq.~\eqref{DiracEq1}. Firstly, the operator $\mathbb{D}$ is symmetric with respect to the inner product~\eqref{innerp2}, \emph{i.e.}, 
\begin{align}
\inpr{\Phi_{\omega_{1}}}{\mathbb{D}\,\Phi_{\omega_2}}=\inpr{\mathbb{D}\,\Phi_{\omega_{1}}}{\Phi_{\omega_2}}\,,
\end{align}
on the domain $\dom{\mathbb{D}}=C_{c}^{\infty}\left((-\pi/2,\pi/2),\mathbb{C}^{2}\right)$, that is, the set of compactly supported smooth spatial spinors with support away from the boundary. The domain of the adjoint operator is found to satisfy
\begin{align*}
\dom{\mathbb{D}}\subseteq\domd{\mathbb{D}}\subseteq AC\left([-\pi/2,\pi/2],\mathbb{C}^2\right)\,,
\end{align*}
where $AC\left(I,\mathbb{C}^2\right)$ is the set of $\mathbb{C}^{2}$--valued functions whose components are absolutely continuous~\cite{reed} on the interval $I\subseteq\mathbb{R}$. In general, $\dom{\mathbb{D}}\neq\domd{\mathbb{D}}$, and thus, the operator $\mathbb{D}$ is not self--adjoint. 

We will therefore look for extensions $\mathbb{D}_U$ of the operator $\mathbb{D}$ that result in self--adjoint operators. To see when these extensions exist, we will apply von~Neumann's theorem~\cite{neu,weyl,reed} as stated in the following form:
\begin{theorem*}[von~Neumann]\label{vNthm}
Let $\mathbb{D}$ be a densely defined symmetric operator on a Hilbert space. Consider the \emph{deficiency subspaces} of $\mathbb{D}$ defined by $\mathscr{K}_{\pm}:=\mathrm{Ker}(\mathbb{D}^{\dagger}\mp i\mathbb{I})$, and the \emph{deficiency indices} of  $\mathbb{D}$ defined as $n_\pm:=\mathrm{dim}(\mathscr{K}_\pm)$. Then,
\begin{enumerate}
\item $\mathbb{D}$ is essentially self--adjoint if and only if $n_{\pm}=0$,
\item $\mathbb{D}$ has self--adjoint extensions if and only if $n_{+}=n_{-}$. The self--adjoint extensions are parametrized by all the isometries from $\mathscr{K}_{+}$ to $\mathscr{K}_{-}$, the correspondence being one--to--one.
\end{enumerate}
\end{theorem*}

To apply this theorem to the operator $\mathbb{D}$ in Eq.~\eqref{OpD} we need to find its deficiency indices $n_{\pm}$. This is equivalent to finding solutions in $L^{2}\left([-\pi/2,\pi/2],\mathbb{C}^{2}\right)$ of the equations $\mathbb{D}^{\dagger}\Phi=\pm i \Phi$, and we note that these equations are of the form of Eq.~\eqref{SpatialDirac1} with $\omega=\pm i$. Therefore,  in order to determine if square--integrable solutions exist for this case, we need to find for which values of $M$ the functions in Eqs.~\eqref{SolComp1} and~\eqref{SolComp2} are square--integrable. This can be done by analyzing the asymptotic behavior of these solutions at the boundary. The leading behavior of the hypergeometric functions appearing in these solutions at $\rho=\pm\pi/2$ is different for different values of the mass of the spinor field $M$ so it will be convenient to perform this analysis separately for the following cases:
\begin{enumerate}[label=(\roman*)]
\item $0\leq M<1/2$.\label{case1}
\item $M>1/2$, with $M-1/2\notin\mathbb{N}$.\label{case2}
\item $M=1/2+k$, with $k\in\mathbb{N}_{0}$.\label{case3}
\end{enumerate}

We analyze cases \ref{case1} and \ref{case2} first. If $M=0$, then it is clear from Eq.~\eqref{MasslessComp1} that both solutions are square--integrable for any $\tilde{C}_1,\tilde{C}_2,\omega\in\mathbb{C}$, therefore, the deficiency indices for the massless case are given by $n_{\pm}=2$. Now, for the non--zero values of $M$ falling on these ranges, we evaluate the functions~\eqref{SolComp1} and~\eqref{SolComp2} at $\rho=\pi/2-\epsilon$ for sufficiently small $\epsilon>0$. Then, using the fact that $F(a,b;c;\epsilon)=1+O(\epsilon)$ as $\epsilon\to 0$, it follows that
\begin{subequations}\label{Asymp-LargeMplus}
\begin{align}
\Phi_{\omega}^{(1)}\left(\frac{\pi}{2}-\epsilon\right)&= \left[(2M+1)C_1+O\left(\epsilon^2\right)\right]\epsilon^{M}+\left[\omega\,C_2+O\left(\epsilon^2\right)\right]\epsilon^{1-M}\,,\label{Asymp-LargeMplus1}\\
\Phi_{\omega}^{(2)}\left(\frac{\pi}{2}-\epsilon\right)&=\left[\omega\,C_1+O\left(\epsilon^2\right)\right]\epsilon^{1+M}+\left[(2M-1)C_2+O\left(\epsilon^2\right)\right]\epsilon^{-M}\,.\label{Asymp-LargeMplus2}
\end{align}
\end{subequations}
To evaluate the component functions near $\rho=-\pi/2$, we use the transformation formula for the hypergeometric function,~\cite{nist}
\begin{align}\label{hyper-transf}
F(a,b;c;x)=&\frac{\Gamma(c)\Gamma(c-a-b)}{\Gamma(c-a)\Gamma(c-b)}F(a,b;a+b-c+1;1-x)\nonumber\\
&+\frac{\Gamma(c)\Gamma(a+b-c)}{\Gamma(a)\Gamma(b)}(1-x)^{c-a-b}F(c-a,c-b;c-a-b+1;1-x)\,,
\end{align}
so that we can write Eqs.~\eqref{SolComp1} and~\eqref{SolComp2} as
\begin{align}\label{SolCompTrans1}
\Phi_{\omega}^{(1)}(\rho)&=(2M+1)C_1\sigma(\rho)^{M}\left[A_{1}^{(M)}F_1^{M}(\rho)+A_{2}^{(M)}\left(1+\sin\rho\right)^{M+\frac{1}{2}}F_2^{M}(\rho)\right]\nonumber\\
&+\omega C_2\cos\rho\,\sigma(\rho)^{-M}\left[B_{1}^{(M)}F_3^{M}(\rho)+B_{2}^{(M)}(1+\sin\rho)^{-M-\frac{1}{2}}F_4^{M}(\rho)\right]\,,
\end{align}
and
\begin{align}\label{SolCompTrans2}
\Phi_{\omega}^{(2)}(\rho)&=\omega\,C_1\cos\rho\,\sigma(\rho)^{-M}\left[B_1^{(-M)}F_3^{-M}(\rho)+B_2^{(-M)}\left(1+\sin\rho\right)^{M-\frac{1}{2}}F_4^{-M}(\rho)\right]\nonumber\\
&+(2M-1)C_2\sigma(\rho)^{-M}\left[A_1^{(-M)}F_1^{-M}(\rho)+A_2^{(-M)}\left(1+\sin\rho\right)^{\frac{1}{2}-M}F_2^{-M}(\rho)\right]\,,
\end{align}
respectively, where we have defined the quantities
\begin{subequations}\label{Trans-Coeffs}
\begin{align}
&A_1^{(M)}=\frac{\Gamma\left(\frac{1}{2}+M\right)^2}{\Gamma\left(\frac{1}{2}+M+\omega\right)\Gamma\left(\frac{1}{2}+M-\omega\right)}\,,\hspace{.25cm}A_2^{(M)}=\frac{\Gamma\left(\frac{1}{2}+M\right)\Gamma\left(-\frac{1}{2}-M\right)}{\Gamma\left(\omega\right)\Gamma\left(-\omega\right)}\,,\\
&B_1^{(M)}=\frac{\Gamma\left(\frac{3}{2}-M\right)\Gamma\left(-M-\frac{1}{2}\right)}{\Gamma\left(\frac{1}{2}-M+\omega\right)\Gamma\left(\frac{1}{2}-M-\omega\right)}\,,\hspace{.25cm}B_2^{(M)}=\frac{\Gamma\left(\frac{3}{2}-M\right)\Gamma\left(\frac{1}{2}+M\right)}{\Gamma\left(1+\omega\right)\Gamma\left(1-\omega\right)}\,.
\end{align}
\end{subequations}
and the functions $F_{j}^{M}(\rho)$, $j=1,\dots 4$, are the hypergeometric functions of argument $(1+\sin\rho)/2$ that result from the transformations in Eq.~\eqref{hyper-transf} and satisfy $F_{j}^{M}(\rho)=1+O(1+\sin\rho)$ as $x\to -\pi/2$. We are now able to evaluate these functions at $\rho=\epsilon-\pi/2$, for the same small parameter $\epsilon>0$ above. This results in
\begin{subequations}\label{Asymp-LargeMminus}
\begin{align}
\Phi_{\omega}^{(1)}\left(\epsilon-\frac{\pi}{2}\right)=&\left[(2M+1)C_1A_1^{(M)}
+C_2\omega\,B_2^{(M)}+O\left(\epsilon^2\right)\right]\epsilon^{-M}\nonumber\\
&+\left[(2M+1)C_1A_2^{(M)}+C_2\omega\,B_1^{(M)}+O\left(\epsilon^2\right)\right]\epsilon^{M+1}\,,\label{Asymp-LargeMminus1}\\
\Phi_{\omega}^{(2)}\left(\epsilon-\frac{\pi}{2}\right)=&\left[C_1\,\omega\,B_1^{(-M)}+(2M-1)C_2A_2^{(-M)}+O\left(\epsilon^2\right)\right]\epsilon^{1-M}\,,\nonumber\\
&+\left[C_1\,\omega\,B_2^{(-M)}+(2M-1)C_2A_1^{(-M)}O\left(\epsilon^2\right)\right]\epsilon^{M}\,.\label{Asymp-LargeMminus2}
\end{align}
\end{subequations}
By the definition of the inner product in Eq.~\eqref{innerp2} it follows that the spinor $\Phi_{\omega}$ will be normalizable if the function $|\Phi_{\omega}^{(1)}(\rho)|^2+|\Phi_{\omega}^{(2)}(\rho)|^2$ goes to $0$ as $\rho\to\pm\pi/2$ sufficiently fast. Using Eq.~\eqref{Asymp-LargeMplus}  and the fact that $M\geq 0$ for the cases we are considering, we have that 
\begin{align}\label{SquareAsympplus}
\left|\Phi_{\omega}^{(1)}\left(\frac{\pi}{2}-\epsilon\right)\right|^{2}+\left|\Phi_{\omega}^{(2)}\left(\frac{\pi}{2}-\epsilon\right)\right|^{2}\sim |C_2|^2\left(|\omega|^2 \epsilon^{2-2M}+(2M-1)^2\epsilon^{-2M}\right)\,,
\end{align}
and,
\begin{align}\label{SquareAsympminus}
\left|\Phi_{\omega}^{(1)}\left(\epsilon-\frac{\pi}{2}\right)\right|^{2}+\left|\Phi_{\omega}^{(2)}\left(\epsilon-\frac{\pi}{2}\right)\right|^{2}\sim &\left|(2M+1)C_1A_1^{(M)}+C_2\,\omega\,B_2^{(M)}\right|^2\epsilon^{-2M}\,,\nonumber\\
&+\left|C_1\,\omega\,B_1^{(-M)}+(2M-1)C_2A_2^{(-M)}\right|^2\epsilon^{2-2M}\,,
\end{align}
as $\epsilon\to 0$. Thus, it follows that if $0< M< 1/2$, then the leading term of these expressions is proportional to $\epsilon^{r}$, with $r>-1$ at both endpoints. Therefore, the two component functions $\Phi_{\omega}^{(1)}$ and $\Phi_{\omega}^{(2)}$ are square--integrable for any $C_1,C_2,\omega\in\mathbb{C}$, in particular, for $\omega=\pm i$, and thus, the deficiency subspaces of $\mathbb{D}$ for this mass range have dimension $n_{\pm}=2$. By von~Neumann's theorem, the operator $\mathbb{D}$ admits a family of self--adjoint extensions parametrized by the isometries from $\mathscr{K}_{+}$ to $\mathscr{K}_{-}$ which, due to finite--dimensionality, can be realized as $2\times 2$--unitary matrices. The self--adjoint extensions of the operator $\mathbb{D}$ for this case will be obtained in Sec.~\ref{SelfAdjSec}.

On the other hand, for case \ref{case2}, if $1/2<M<3/2$, the terms with $\epsilon^{2-2M}$ decay faster than $\epsilon^{-1}$, so the singular behavior comes from the terms proportional to $\epsilon^{-2M}$. For the spinor $\Phi_{\omega}$ to be square--integrable at both endpoints, we must have
\begin{align}\label{aux1}
C_2=0\,,\hspace{.5cm}\text{and}\hspace{.5cm}A_1^{(M)}=0\,.
\end{align}
If $M>3/2$ and $M-1/2\notin\mathbb{N}$, then terms proportional to $\epsilon^{2-2M}$ are also singular, so square--integrable solutions for this range need to satisfy Eq.~\eqref{aux1} as well as the additional condition
\begin{align}
\omega\,B_1^{(-M)}=0\,.
\end{align}
Using the definitions of these quantities in Eq.~\eqref{Trans-Coeffs} we note that $B_1^{(-M)}$ is proportional to $A_1^{(M)}$ as a function of $\omega$, so the only positive values of $\omega$ for which $A_1^{(M)}$, and therefore $B_1^{(-M)}$, vanish are given by $\omega=\omega_{n}^{I}$, where
\begin{align}\label{frequency}
\omega_{n}^{I}:=\frac{1}{2}+M+n\,,\hspace{.5cm}n\in\mathbb{N}\cup\{0\}\,.
\end{align}
This implies that no square--integrable solutions exist for $\omega=\pm i$, and therefore, the deficiency spaces are both zero--dimensional and thus, by von~Neumann's theorem, the operator $\mathbb{D}$ is essentially self--adjoint. This means that the unique self--adjoint extension for the operator $\mathbb{D}$ is its closure~\cite{reed} $\mathbb{\overline{D}}$.

A similar conclusion holds for case \ref{case3}. First we note that for the values of $M$ we are considering,  the general solution to Eq.~\eqref{CoupDir1} with $\omega=0$ is given by $\Phi(\rho)=C_1(\sigma(\rho)^{M},0)^{T}+C_2(0,\sigma(\rho)^{-M})^{T}$, where $\sigma(\rho)$ is defined by Eq.~\eqref{sol-pref}. This solution is not square--integrable for any $M\geq 1/2$, so we will continue the analysis for these values of $M$ assuming $\omega\neq 0$. Let us now consider the functions in Eq.~\eqref{SolCompklog} with $k>0$. As shown in Appendix~\ref{App1}, the asymptotic behavior of the these component functions at $\rho=\pi/2-\epsilon$, for sufficiently small $\epsilon>0$ is given by
\begin{subequations}\label{SolCompkAsymp-plus}
\begin{align}
\Phi_{\omega}^{(1)}\left(\frac{\pi}{2}-\epsilon\right)&= \left[C_2 \omega A_{3}^{(k)}+O\left(\epsilon^2\right)\right]\epsilon^{-k+\frac{1}{2}}\,,\label{SolCompkAsymp-plus1}\\
\Phi_{\omega}^{(2)}\left(\frac{\pi}{2}-\epsilon\right)&=\left[ C_2 k A_{3}^{(k)}+O\left(\epsilon^2\right)\right] \epsilon^{-k-\frac{1}{2}}\,,\label{SolCompkAsymp-plus2}
\end{align}
\end{subequations}
and similarly, at $\rho=\epsilon-\pi/2$, we have 
\begin{subequations}\label{SolCompkAsymp-minus}
\begin{align}
\Phi_{\omega}^{(1)}\left(\epsilon-\frac{\pi}{2}\right)&= k\,A_3^{(k)}\left[C_1\frac{2}{\pi}\sin\pi(\omega-k)+C_2\cos\pi(\omega-k)+O\left(\epsilon^2\right)\right]\epsilon^{-k-\frac{1}{2}}\,,\label{SolCompkAsymp-minus1}\\
\Phi_{\omega}^{(2)}\left(\epsilon-\frac{\pi}{2}\right)&= \omega\,A_3^{(k)} \left[C_1\frac{2}{\pi}\sin\pi(\omega-k)+C_2\cos\pi(\omega-k)+O\left(\epsilon^2\right)\right]\epsilon^{-k+\frac{1}{2}}\,,\label{SolCompkAsymp-minus2}
\end{align}
\end{subequations}
where we have defined 
\begin{align}\label{A3coeff}
A_3^{(k)}:=\frac{2^{k}\Gamma(k)\Gamma(\omega-k)}{\Gamma(\omega+k+1)}\,.
\end{align}
The behavior of the modulus squared of the spinor at the boundary is obtained using Eqs.~\eqref{SolCompkAsymp-plus} and~\eqref{SolCompkAsymp-minus} and it can readily be verified that it is given by Eq.~\eqref{SquareAsympplus} at $\rho=\pi/2$ and by Eq.~\eqref{SquareAsympminus} at $\rho=-\pi/2$, with $M=k+1/2$. From these approximations it is clear that the leading terms at both endpoints are of the form $\epsilon^{r}$ with $r<-1$, so in order to obtain square--integrable solutions the expressions on the left--hand side must vanish simultaneously. This occurs only when $|A_3^{(k)}|^2=0$, or when $C_2=0$ and $\sin\pi(\omega-k)=0$. From Eq.~\eqref{A3coeff}, the former case only happens when $\omega=-n-k-1$  with $n\in\mathbb{N}\cup\{0\}$, and the latter only happens when $\omega=n+k+1$. Therefore, no square--integrable solutions for $\omega=\pm i$ exist for this case either, thus, the deficiency indices are once again $n_\pm =0$, hence, the unique self--adjoint extension is given by the closure $\mathbb{\overline{D}}$. 

If $k=0$, then the solutions in Eq.~\eqref{SolCompklog} are given in terms of Legendre functions which have a different asymptotic expansion at the endpoints of the boundary. From the analysis in Appendix~\ref{App1} we find that 
\begin{subequations}\label{Squarek0Asymp}
\begin{align}
\left|\Phi_{\omega}^{(1)}\left(\frac{\pi}{2}-\epsilon\right)\right|^{2}+\left|\Phi_{\omega}^{(2)}\left(\frac{\pi}{2}-\epsilon\right)\right|^{2}
&\sim \frac{|C_2|^2}{|\omega|^2}\,\epsilon^{-1}\,,\label{Squarek0Asymp1}\\
\left|\Phi_{\omega}^{(1)}\left(\epsilon-\frac{\pi}{2}\right)\right|^{2}+\left|\Phi_{\omega}^{(2)}\left(\epsilon-\frac{\pi}{2}\right)\right|^{2}&\sim\left|C_1\frac{2}{\pi\omega}\sin\pi\omega+C_2\frac{1}{\omega}\cos\pi\omega\right|^2\epsilon^{-1}\,,\label{Squarek0Asymp2}
\end{align}
\end{subequations}
as $\epsilon\to 0$. Once again, the spinor solution will be square--integrable if the above expressions on the left--hand side vanish. This only happens if $C_2=0$ and $\sin\pi\omega=0$, the latter condition restricting the values of $\omega$ to be $\omega\in\mathbb{Z}$, and we once again note that these values are also of the form $\omega_{n}^{(I)}$ in Eq.~\eqref{frequency} with $M=1/2$. Thus, by the same argument as for the case $k>0$, the deficiency spaces are zero--dimensional, and thus we will treat this case in a way similar to the case~\ref{case2}.

Let us summarize the results of this section. Looking at the asymptotic behavior of the solutions at the boundary we have found that the deficiency indices of the operator $\mathbb{D}$ are $n_\pm=2$ when $0\leq M<1/2$ (case \ref{case1} above) and $n_{\pm}=0$ when the mass of the spinor field satisfies $M\geq 1/2$ (cases \ref{case2} and \ref{case3}). For the latter case, square--integrable solutions exist only for the frequencies given by Eq.~\eqref{frequency}. Now that we have found the deficiency indices of the operator $\mathbb{D}$, we will proceed to obtain the associated self--adjoint extensions $\mathbb{D}_U$. 

\subsection{Self-adjoint extensions of the operator $\mathbb{D}$}\label{SelfAdjSec}

In this section we will find the self--adjoint extensions of the operator $\mathbb{D}$. We start by analyzing the case with $0\leq M<1/2$ for which we have found that $n_{\pm}=2$. It is a well known fact~\cite{reed} that for a symmetric operator $\mathbb{D}$ with closure $\mathbb{\overline{D}}$ and deficiency subspaces $\mathscr{K}_{\pm}$, the domain of its adjoint operator $\mathbb{D}^{\dagger}$ is given by
\begin{align}\label{adjoint-domain}
\domd{\mathbb{D}}=\dom{\overline{\mathbb{D}}}\oplus\mathscr{K}_{+}\oplus\mathscr{K}_{-}\,,
\end{align}
and any closed self--adjoint extension $\mathbb{D}_U$ has domain given by
\begin{align}\label{ext-domain}
\dom{\mathbb{D}_U}=\dom{\overline{\mathbb{D}}}\oplus\mathscr{S}\,,
\end{align}
where $\mathscr{S}\subseteq\mathscr{K}_{+}\oplus\mathscr{K}_{-}$ is a maximal subspace on which the operator $\mathbb{D}^{\dagger}$ is symmetric. This fact makes possible the description of all the self--adjoint extensions of the operator $\mathbb{D}$ in terms of boundary conditions which we will impose to the solutions of Eq.~\eqref{DiracEq3}. These boundary conditions are obtained by finding the conditions that elements of $\mathscr{S}$ satisfy. For this purpose we will need to manipulate the boundary values of the solutions with $0\leq M<1/2$ which, by the analysis in Sec.~\ref{DiracEqSec}, were found to be given only in terms of the asymptotic behavior of the solutions at the boundary. Therefore, it will be convenient to define the component functions $\widetilde{\Phi^{(1)}}$ and $\widetilde{\Phi^{(2)}}$ that contain the same leading behavior as the components $\Phi^{(1)}$ and $\Phi^{(2)}$ but take on finite values when evaluated at $\rho=\pm\pi/2$. Hence, given $\Phi\in\domd{\mathbb{D}}$ and $0<M<1/2$, we define
\begin{subequations}\label{WeightedComp}
\begin{align}
\widetilde{\Phi^{(1)}}(\rho)&:=\sigma(\rho)^{-M}\Phi^{(1)}(\rho)\,,\label{WeightedComp1}\\
\widetilde{\Phi^{(2)}}(\rho)&:=\sigma(\rho)^{M}\Phi^{(2)}(\rho)\,.\label{WeightedComp2}
\end{align}
\end{subequations}
For the case $M=0$ the functions on the left--hand side are defined trivially by $\widetilde{\Phi^{(1)}}=\Phi^{(1)}$ and $\widetilde{\Phi^{(2)}}=\Phi^{(2)}$. In this way if $\Phi_{\omega}$ is any solution of Eq.~\eqref{DiracEq3} with $0< M<1/2$ and fixed $\omega\in\mathbb{C}$ then, with Eqs.~\eqref{Asymp-LargeMplus} and~\eqref{Asymp-LargeMminus}, a straightforward calculation shows that the component functions defined through Eq.~\eqref{WeightedComp} satisfy
\begin{subequations}\label{BoundaryValues}
\begin{align}
\begin{pmatrix}
\widetilde{\Phi_{\omega}^{(1)}}\left(\frac{\pi}{2}\right)\\
\widetilde{\Phi_{\omega}^{(2)}}\left(\frac{\pi}{2}\right)
\end{pmatrix}&=\begin{pmatrix}
(2M+1)C_1\\[1em]
(2M-1)C_2
\end{pmatrix}\,,\label{BoundaryValues1}\\
\begin{pmatrix}
\widetilde{\Phi_{\omega}^{(1)}}\left(-\frac{\pi}{2}\right)\\
\widetilde{\Phi_{\omega}^{(2)}}\left(-\frac{\pi}{2}\right)
\end{pmatrix}&=\begin{pmatrix}
(2M+1)C_1A_1^{(M)}+2^{M+\frac{1}{2}}C_2\,\omega\,B_2^{(M)}\\[1em]
2^{M-\frac{1}{2}}C_1\,\omega\,B_2^{(-M)}+(2M-1)C_2\,A_1^{(-M)}
\end{pmatrix}\,,\label{BoundaryValues2}
\end{align}
\end{subequations}
and from Eq.~\eqref{MasslessComp1} it is clear that solutions for $M=0$ satisfy the simpler relations
\begin{align}\label{BoundaryValues0}
\begin{pmatrix}
\widetilde{\Phi_{\omega}^{(1)}}\left(\pm\frac{\pi}{2}\right)\\[.5em]
\widetilde{\Phi_{\omega}^{(2)}}\left(\pm\frac{\pi}{2}\right)
\end{pmatrix}=\begin{pmatrix}
\tilde{C}_1\cos\frac{\omega\pi}{2}\pm \tilde{C}_2\sin\frac{\omega\pi}{2}\\[.5em]
\mp \tilde{C}_1\sin\frac{\omega\pi}{2}+\tilde{C}_2\cos\frac{\omega\pi}{2}
\end{pmatrix}\,.
\end{align}

Let us recall that the elements of deficiency subspaces $\mathscr{K}_{+}$ and $\mathscr{K}_{-}$ are linear combinations of the solutions $\Phi_{\omega}$ of Eq.~\eqref{SpatialDirac1} with $\omega=\pm i$, respectively. If $M=0$, a solution $\Phi_{\pm i}$ of this equation is given by Eq.~\eqref{MasslessComp1} and, if $M\neq 0$ it is given by Eqs.~\eqref{SolComp1} and~\eqref{SolComp2} instead. Since  $\mathscr{K}_{\pm}$ is two dimensional, its elements are characterized by the two coefficients of the solution $\Phi_{\pm i}$ which we can denote by $C_{1}^{\pm}$ and $C_{2}^{\pm}$. Equations~\eqref{BoundaryValues} and~\eqref{BoundaryValues0}, with $C_1,C_2,\tilde{C}_{1},\tilde{C}_{2}$ replaced by $C_{1}^{\pm}$ and $C_{2}^{\pm}$, imply that these coefficients are completely determined by the values of the functions $\widetilde{\Phi_{\pm i}^{(1)}}$ and $\widetilde{\Phi_{\pm i}^{(2)}}$ at the boundary. Therefore, an element $\Phi=\Phi_{i}+\Phi_{-i}\in\mathscr{K}_{+}\oplus\mathscr{K}_{-}$ is in one--to--one correspondence with the vector
\begin{align}\label{BoundDataVec}
\left(\widetilde{\Phi^{(1)}}\left(\frac{\pi}{2}\right),\widetilde{\Phi^{(2)}}\left(\frac{\pi}{2}\right),\widetilde{\Phi^{(1)}}\left(-\frac{\pi}{2}\right),\widetilde{\Phi^{(2)}}\left(-\frac{\pi}{2}\right)\right)^{T}\in\mathbb{C}^{4}\,,
\end{align}
of boundary data. This means that finding the two--dimensional subspace $\mathscr{S}\subset\mathscr{K}_{+}\oplus\mathscr{K}_{-}$ that characterizes a self--adjoint extension of $\mathbb{D}$ is equivalent to finding the two dimensional subspace of $\mathbb{C}^{4}$ of boundary data on which $\mathbb{D}^{\dagger}$ is symmetric. Now, from the definition of the subspace $\mathscr{S}\subset\mathscr{K}_{+}\oplus\mathscr{K}_{-}$, any element $\Phi\in\mathscr{S}$ must satisfy
\begin{align}
0=\inpr{\mathbb{D}^{\dagger}\Phi}{\Phi}-\inpr{\Phi}{\mathbb{D}^{\dagger}\Phi}\,.
\end{align}
Using the inner product from Eq.~\eqref{innerp2}, expanding in terms of the components $\Phi^{(1)}$ and $\Phi^{(2)}$ of the spinor $\Phi$ and integrating by parts, this condition becomes
\begin{align}\label{Boundary-Cond2}
0&=\left.\left[\overline{\Phi^{(1)}(\rho)}\Phi^{(2)}(\rho)-\overline{\Phi^{(2)}(\rho)}\Phi^{(1)}(\rho)\right]\right|_{-\pi/2}^{\pi/2}\,,\nonumber\\
&=\left.\left[\overline{\widetilde{\Phi^{(1)}}(\rho)}\widetilde{\Phi^{(2)}}(\rho)-\overline{\widetilde{\Phi^{(2)}}(\rho)}\widetilde{\Phi^{(1)}}(\rho)\right]\right|_{-\pi/2}^{\pi/2}\,,
\end{align}
where the second equality is obtained  using the definition in Eq.~\eqref{WeightedComp}. We can rewrite Eq.~\eqref{Boundary-Cond2} as
\begin{align}
0=& \left|\widetilde{\Phi^{(2)}}\left(\frac{\pi}{2}\right)+i\widetilde{\Phi^{(1)}}\left(\frac{\pi}{2}\right)\right|^{2}-\left|\widetilde{\Phi^{(2)}}\left(\frac{\pi}{2}\right)-i\widetilde{\Phi^{(1)}}\left(\frac{\pi}{2}\right)\right|^{2}\nonumber\\
&+\left|\widetilde{\Phi^{(2)}}\left(-\frac{\pi}{2}\right)-i\widetilde{\Phi^{(1)}}\left(-\frac{\pi}{2}\right)\right|^{2}-\left|\widetilde{\Phi^{(2)}}\left(-\frac{\pi}{2}\right)+i\widetilde{\Phi^{(1)}}\left(-\frac{\pi}{2}\right)\right|^{2}\,.
\end{align}
This expression implies that the components of the vectors of the form of Eq.~\eqref{BoundDataVec} that belong to the symmetric subspace $\mathscr{S}$ must satisfy
\begin{align}\label{BoundAuxEq1}
\left|\begin{pmatrix}
\widetilde{\Phi^{(2)}}\left(\frac{\pi}{2}\right)+i\widetilde{\Phi^{(1)}}\left(\frac{\pi}{2}\right)\\
\widetilde{\Phi^{(2)}}\left(-\frac{\pi}{2}\right)-i\widetilde{\Phi^{(1)}}\left(-\frac{\pi}{2}\right)
\end{pmatrix}\right|^{2}=\left|\begin{pmatrix}
\widetilde{\Phi^{(2)}}\left(\frac{\pi}{2}\right)-i\widetilde{\Phi^{(1)}}\left(\frac{\pi}{2}\right)\\
\widetilde{\Phi^{(2)}}\left(-\frac{\pi}{2}\right)+i\widetilde{\Phi^{(1)}}\left(-\frac{\pi}{2}\right)
\end{pmatrix}\right|^{2}\,.
\end{align}
By writing the element from Eq.~\eqref{BoundDataVec} as $(\boldsymbol{\phi}_{1},\boldsymbol{\phi}_{2})^{T}$, where
\begin{align}\label{phivec}
\boldsymbol{\phi}_{1}=\begin{pmatrix}
\widetilde{\Phi^{(2)}}\left(\frac{\pi}{2}\right)+i\widetilde{\Phi^{(1)}}\left(\frac{\pi}{2}\right)\\
\widetilde{\Phi^{(2)}}\left(-\frac{\pi}{2}\right)-i\widetilde{\Phi^{(1)}}\left(-\frac{\pi}{2}\right)
\end{pmatrix}\,,\hspace{.5cm}\boldsymbol{\phi}_{2}=\begin{pmatrix}
\widetilde{\Phi^{(2)}}\left(\frac{\pi}{2}\right)-i\widetilde{\Phi^{(1)}}\left(\frac{\pi}{2}\right)\\
\widetilde{\Phi^{(2)}}\left(-\frac{\pi}{2}\right)+i\widetilde{\Phi^{(1)}}\left(-\frac{\pi}{2}\right)
\end{pmatrix}\,,
\end{align}
Eq.~\eqref{BoundAuxEq1} together with the linearity of $\mathscr{S}$ imply that if $(\boldsymbol{0},\boldsymbol{\phi}_2)^{T}\in\mathscr{S}$, then $\boldsymbol{\phi}_{2}=\boldsymbol{0}$, and thus, if $(\boldsymbol{\phi}_{1},\boldsymbol{\phi}_{2})^{T},(\boldsymbol{\phi}_{1},\boldsymbol{\phi'}_{2})^{T}\in\mathscr{S}$, then $\boldsymbol{\phi}_{2}=\boldsymbol{\phi'}_{2}$. Therefore, any $(\boldsymbol{\phi}_{1},\boldsymbol{\phi}_{2})^{T}\in\mathscr{S}$ must satisfy $\boldsymbol{\phi_2}=U(\boldsymbol{\phi_{1}})$ for some linear function $U$. Hence, for any $\Phi\in\mathscr{S}$, we must have
\begin{align}\label{SABCzero}
U\begin{pmatrix}
\widetilde{\Phi^{(2)}}\left(\frac{\pi}{2}\right)+i\widetilde{\Phi^{(1)}}\left(\frac{\pi}{2}\right)\\
\widetilde{\Phi^{(2)}}\left(-\frac{\pi}{2}\right)-i\widetilde{\Phi^{(1)}}\left(-\frac{\pi}{2}\right)
\end{pmatrix}=\begin{pmatrix}
\widetilde{\Phi^{(2)}}\left(\frac{\pi}{2}\right)-i\widetilde{\Phi^{(1)}}\left(\frac{\pi}{2}\right)\\
\widetilde{\Phi^{(2)}}\left(-\frac{\pi}{2}\right)+i\widetilde{\Phi^{(1)}}\left(-\frac{\pi}{2}\right)
\end{pmatrix}\,,
\end{align}
where $U$ is a $2\times 2$ matrix. For the boundary values of $\Phi\in\mathscr{S}$ to span a two dimensional space under this condition, the vector $\boldsymbol{\phi}_1$ of Eq.~\eqref{phivec} must take on all possible values. This fact together with Eq.~\eqref{BoundAuxEq1} imply that $U$ must be a unitary matrix. Conversely, for any two $\Phi,\Phi'\in\mathscr{K}_{+}\oplus\mathscr{K}_{-}$ satisfying Eq.~\eqref{SABCzero} for a fixed unitary matrix $U$, we have $\inpr{\mathbb{D}^{\dagger}\Phi}{\Phi'}=\inpr{\Phi}{\mathbb{D}^{\dagger}\Phi'}$. Thus, we conclude that the unitary matrix $U$ specifies a self--adjoint extension $\mathbb{D}_{U}$ of the operator $\mathbb{D}$ through the boundary data of the element $\Phi$. In this way, every self--adjoint extension $\mathbb{D}_{U}$ is characterized by the boundary condition~\eqref{SABCzero} which we rewrite as
\begin{align}\label{SABC1}
(\mathbb{I}-U)\begin{pmatrix}
\widetilde{\Phi^{(2)}}\left(\frac{\pi}{2}\right)\\
\widetilde{\Phi^{(2)}}\left(-\frac{\pi}{2}\right)
\end{pmatrix}=i(\mathbb{I}+U)\begin{pmatrix}
\widetilde{\Phi^{(1)}}\left(\frac{\pi}{2}\right)\\
-\widetilde{\Phi^{(1)}}\left(-\frac{\pi}{2}\right)
\end{pmatrix}\,.
\end{align}
With this, we are now able to solve the eigenvalue problem of Eq.~\eqref{DiracEq3}. We will solve the equation $\mathbb{D}^{\dagger}\Phi_{\omega}=\omega\Phi_{\omega}$ now with $\omega\in\mathbb{R}$ to be determined by requiring that the solutions as given by Eqs.~\eqref{SolComp1} and~\eqref{SolComp2} are elements of $\dom{\mathbb{D}_U}$. By the analysis carried out in this section, this is equivalent to find solutions to this equation satisfying the boundary condition in Eq.~\eqref{SABC1} for some matrix $U$. 

Finally, for the solutions with mass satisfying $M\geq 1/2$, as shown in Sec.~\ref{DiracEqSec}, the deficiency indices satisfy $n_{\pm}=0$ and thus, the only self--adjoint extension of the operator $\mathbb{D}$ is its closure $\mathbb{\overline{D}}$. From Eq.~\eqref{adjoint-domain}, we see that $\dom{\overline{\mathbb{D}}}=\domd{\mathbb{D}}$, so finding solutions in $\dom{\overline{\mathbb{D}}}$ is equivalent to finding square--integrable solutions of $\mathbb{D}^{\dagger}\Phi_{\omega}=\omega\Phi_{\omega}$. We recall that, from the analysis of the deficiency spaces in Sec.~\ref{DiracEqSec}, square--integrable solutions for this equation exist only when the frequency $\omega$ is restricted to be of the form $\omega_{n}^{I}$ as given in Eq.~\eqref{frequency}. We can rephrase this fact in a more similar way to that of the cases with $0\leq M</2$, that is, in terms of a  boundary condition as follows: From the definitions of the component functions $\widetilde{\Phi_{\omega}^{(1)}}$ and $\widetilde{\Phi_{\omega}^{(2)}}$ in Eq.~\eqref{WeightedComp}, it can be directly verified that imposing the boundary condition, which we will refer to as \emph{Dirichlet type I} boundary condition, given by
\begin{align}\label{aux2}
\widetilde{\Phi_{\omega}^{(2)}}\left(\frac{\pi}{2}\right)=0=\widetilde{\Phi_{\omega}^{(1)}}\left(-\frac{\pi}{2}\right)\,,
\end{align}
on the general solutions in Eqs.~\eqref{SolComp1} and~\eqref{SolComp2} if $M-1/2\notin\mathbb{N}\cup\{0\}$ and, in Eq.~\eqref{SolCompklog} if $M-1/2\in\mathbb{N}\cup\{0\}$, results in the same frequency spectrum $\omega_{n}^{I}$ and restriction $C_2=0$ that were obtained in Sec.~\ref{DiracEqSec} by requiring square--integrability of the spatial spinor $\Phi_{\omega}$. Hence, we can conclude that for all $M\geq 1/2$, the unique self--adjoint extension of $\mathbb{D}$ is characterized by the Dirichlet type I boundary condition. This fact will be used in Sec.~\ref{InvariantBC-Sec} when analyzing the invariance of the solutions under the action of $\SLC$.

\subsection{Invariant self-adjoint boundary conditions}\label{InvariantBC-Sec}

Instead of finding the spectrum of the operator $\mathbb{D}_U$ for every unitary matrix $U$, we will only focus on those matrices which, via Eq.~\eqref{SABC1}, result in boundary conditions that remain invariant under infinitesimal $\SLC$--transformations.

The infinitesimal action of $\SLC$ on spinors is realized thought the operators $\mathcal{L}_\pm$ defined in Eq.~\eqref{ladderop1}. Recalling that we have been considering spinor solutions of the form given by Eq.~\eqref{prefactor1}, we will denote by $L_\pm$ the associated ladder operators acting on the spinor $\Psi=(\cos\rho)^{-1/2}\psi$. Following the decomposition into mode solutions from Eq.~\eqref{modedecomp1}, it follows that the action of the ladder operators on spinors of the form $\Psi(t,\rho)=\Phi_{\omega}(\rho)e^{-i\omega t}$ is given by
\begin{align}\label{Ladderaction}
L_\pm\left[\Phi_{\omega}(\rho)e^{-i\omega t}\right]&=\pm i\left[\cos\rho\,\frac{\D}{\D\rho}-\left(\frac{1}{2}\pm\omega\right)\sin\rho\mp i\cos\rho\,\Sigma^{01}\right]\Phi_{\omega}(\rho)e^{-i(\omega\pm 1)t}\,.
\end{align}
From this expression we see that at $t=0$, the component functions $\Phi_{\omega}^{(1)}$ and $\Phi_{\omega}^{(2)}$ of the spinor $\Phi_\omega$ transform under the action of $\mp iL_{\pm}$ as
\begin{subequations}\label{Transfcomp-noDer}
\begin{align}
(\delta_{\pm}\Phi_{\omega})^{(1)}(\rho)&:=-\left[M+\left(\frac{1}{2}\pm\omega\right)\sin\rho\right]\Phi_{\omega}^{(1)}(\rho)+\left(\omega\pm\frac{1}{2}\right)\cos\rho\,\Phi_{\omega}^{(2)}(\rho)\,,\label{Transfcomp-noDer1}\\
(\delta_{\pm}\Phi_{\omega})^{(2)}(\rho)&:=-\left(\omega\pm\frac{1}{2}\right)\cos\rho\,\Phi_{\omega}^{(1)}(\rho)+\left[M-\left(\frac{1}{2}\pm\omega\right)\sin\rho\right]\Phi_{\omega}^{(2)}(\rho)\,.\label{Transfcomp-noDer1}
\end{align}
\end{subequations}
where we have used the spatial Dirac equation in the form of Eq. \eqref{CoupDir1} to eliminate the derivative terms. Using the definitions of the components $\widetilde{\Phi_{\omega}^{(1)}}$ and $\widetilde{\Phi_{\omega}^{(2)}}$ in Eq.~\eqref{WeightedComp}, we find that 
\begin{subequations}\label{Transf-TildComp}
\begin{align}
\widetilde{(\delta_{\pm}\Phi_{\omega})^{(1)}}(\rho)&=-\left[M+\left(\frac{1}{2}\pm\omega\right)\sin\rho\right]\widetilde{\Phi_{\omega}^{(1)}}(\rho)+\left(\omega\pm\frac{1}{2}\right)\cos\rho\,\sigma(\rho)^{-2M}\widetilde{\Phi_{\omega}^{(2)}}(\rho)\,,\label{Transf-TildComp1}\\
\widetilde{(\delta_{\pm}\Phi_{\omega})^{(2)}}(\rho)&=-\left(\omega\pm\frac{1}{2}\right)\cos\rho\,\sigma(\rho)^{2M}\widetilde{\Phi_{\omega}^{(1)}}(\rho)+\left[M-\left(\frac{1}{2}\pm\omega\right)\sin\rho\right]\widetilde{\Phi_{\omega}^{(2)}}(\rho)\,.\label{Transf-TildComp2}
\end{align}
\end{subequations}
In particular, for the $\delta_{-}$--transformation, we have
\begin{subequations}\label{TildeBoundaryValues}
\begin{align}
\widetilde{(\delta_{-}\Phi_{\omega})^{(1)}}\left(\pm\frac{\pi}{2}\right)&=-\left[M\pm\left(\frac{1}{2}-\omega\right)\right]\widetilde{\Phi_{\omega}^{(1)}}\left(\pm\frac{\pi}{2}\right)\,,\\
\widetilde{(\delta_{-}\Phi_{\omega})^{(2)}}\left(\pm\frac{\pi}{2}\right)&=\left[M\mp\left(\frac{1}{2}-\omega\right)\right]\widetilde{\Phi_{\omega}^{(2)}}\left(\pm\frac{\pi}{2}\right)\,.
\end{align}
\end{subequations}
and the values for $\delta_{+}$ are obtained by replacing $\omega\mapsto-\omega$. We also note that these boundary values are valid for the massless case by setting $M=0$.

Now, if  $\Phi_{\omega_{1}}$ and $\Phi_{\omega_{2}}$ are two solutions of the equation $\mathbb{D}_U\Phi=\omega\Phi$, with $\omega_1,\omega_2\in\mathbb{R}$, satisfying the same boundary condition~\eqref{SABC1} for a fixed matrix $U$, then we must have $0=\inpr{\mathbb{D}_U\Phi_{\omega_1}}{\Phi_{\omega_{2}}}-\inpr{\Phi_{\omega_1}}{\mathbb{D}_U\Phi_{\omega_{2}}}$, this is,
\begin{align}\label{BCondition1}
0=&\left[\overline{\widetilde{\Phi_{\omega_1}^{(1)}}\left(\frac{\pi}{2}\right)}\widetilde{\Phi_{\omega_2}^{(2)}}\left(\frac{\pi}{2}\right)-\overline{\widetilde{\Phi_{\omega_1}^{(2)}}\left(\frac{\pi}{2}\right)}\widetilde{\Phi_{\omega_2}^{(1)}}\left(\frac{\pi}{2}\right)\right]\nonumber\\
&-\left[\overline{\widetilde{\Phi_{\omega_1}^{(1)}}\left(-\frac{\pi}{2}\right)}\widetilde{\Phi_{\omega_2}^{(2)}}\left(-\frac{\pi}{2}\right)-\overline{\widetilde{\Phi_{\omega_1}^{(2)}}\left(-\frac{\pi}{2}\right)}\widetilde{\Phi_{\omega_2}^{(1)}}\left(-\frac{\pi}{2}\right)\right]\,.
\end{align}
Using Eq.~\eqref{TildeBoundaryValues}, it follows that if the boundary condition is invariant under the transformation induced by $L_\pm$, then the relation
\begin{align}\label{BCondition2}
0=&\delta_{-}\left[\inpr{\mathbb{D}_U\Phi_{\omega_1}}{\Phi_{\omega_{2}}}-\inpr{\Phi_{\omega_1}}{\mathbb{D}_U\Phi_{\omega_{2}}}\right]\,,\nonumber\\
=&-(1-\omega_1-\omega_2)\left[\overline{\widetilde{\Phi_{\omega_1}^{(1)}}\left(\frac{\pi}{2}\right)}\widetilde{\Phi_{\omega_2}^{(2)}}\left(\frac{\pi}{2}\right)-\overline{\widetilde{\Phi_{\omega_1}^{(2)}}\left(\frac{\pi}{2}\right)}\widetilde{\Phi_{\omega_2}^{(1)}}\left(\frac{\pi}{2}\right)\right]\nonumber\\
&-(1-\omega_1-\omega_2)\left[\overline{\widetilde{\Phi_{\omega_1}^{(1)}}\left(-\frac{\pi}{2}\right)}\widetilde{\Phi_{\omega_2}^{(2)}}\left(-\frac{\pi}{2}\right)-\overline{\widetilde{\Phi_{\omega_1}^{(2)}}\left(-\frac{\pi}{2}\right)}\widetilde{\Phi_{\omega_2}^{(1)}}\left(-\frac{\pi}{2}\right)\right]\,,
\end{align}
(here the operator $\mathbb{D}_U$ is not transformed) must also be satisfied. Eqs.~\eqref{BCondition1} and~\eqref{BCondition2} are compatible with each other if and only if the two equations
\begin{align}
\left[\overline{\widetilde{\Phi_{\omega_1}^{(1)}}\left(\pm\frac{\pi}{2}\right)}\widetilde{\Phi_{\omega_2}^{(2)}}\left(\pm\frac{\pi}{2}\right)-\overline{\widetilde{\Phi_{\omega_1}^{(2)}}\left(\pm\frac{\pi}{2}\right)}\widetilde{\Phi_{\omega_2}^{(1)}}\left(\pm\frac{\pi}{2}\right)\right]=0\,,
\end{align}
are simultaneously satisfied. For this to hold for all pairs $\{\omega_1,\omega_2\}$,
\begin{align}
\widetilde{\Phi_{\omega}^{(2)}}\left(\pm\frac{\pi}{2}\right)\propto\widetilde{\Phi_{\omega}^{(1)}}\left(\pm\frac{\pi}{2}\right)\,,
\end{align}
for both $\omega_1$ and $\omega_2$, with the proportionality constant being the same real number, or $\widetilde{\Phi_{\omega}^{(1)}}=0$ at each endpoint. This is only true if the unitary matrix $U$ in Eq.~\eqref{SABC1} is diagonal, that is, if
\begin{align}\label{Diagbc}
(1-e^{i\alpha_{\pm}})\widetilde{\Phi_{\omega}^{(2)}}\left(\pm\frac{\pi}{2}\right)=\pm i(1+e^{i\alpha_{\pm}})\widetilde{\Phi_{\omega}^{(1)}}\left(\pm\frac{\pi}{2}\right)\,,
\end{align}
for some $\alpha_\pm\in\mathbb{R}$.

Now, using Eq.~\eqref{Transf-TildComp} we see that that if these boundary conditions are invariant, then the components $\left(\delta_{-}\Phi_{\omega}\right)^{(1)}$ and $\left(\delta_{-}\Phi_{\omega}\right)^{(2)}$  must satisfy
\begin{align}\label{Diagbctransf}
\left(\frac{1}{2}-\omega\mp M\right)(1-e^{i\alpha_{\pm}})\widetilde{\Phi_{\omega}^{(2)}}\left(\pm\frac{\pi}{2}\right)=\pm i\left(\frac{1}{2}-\omega\pm M\right)(1+e^{i\alpha_{\pm}})\widetilde{\Phi_{\omega}^{(1)}}\left(\pm\frac{\pi}{2}\right)\,.
\end{align}
From these relations, it is clear that if $M=0$, then Eq.~\eqref{Diagbctransf} reduces to Eq.~\eqref{Diagbc}, so the self--adjoint extensions parametrized by a diagonal matrix $U$ are all invariant under both ladder operators. In contrast, if $0<M<1/2$, then for Eq.~\eqref{Diagbc} to be consistent with Eq.~\eqref{Diagbctransf}, there are only 4 different cases for the values that the numbers $e^{i\alpha_\pm}$ can take:
\begin{enumerate}
\item $e^{i\alpha_\pm}=\mp 1$, \emph{i.e.}, $U=\mathrm{diag}(-1,1)$. Then the invariant boundary condition is the Dirichlet type I condition that were also found for the case $M\geq 1/2$ in Eq.~\eqref{aux2}, namely,
\begin{align}\label{Bc1}
\widetilde{\Phi_{\omega}^{(2)}}\left(\frac{\pi}{2}\right)=0=\widetilde{\Phi_{\omega}^{(1)}}\left(-\frac{\pi}{2}\right)\,.
\end{align}
\item $e^{i\alpha_{\pm}}=\pm 1$, \emph{i.e.}, $U=\mathrm{diag}(1,-1)$. Then, Eq.~\eqref{SABC1} reduces to
\begin{align}\label{Bc2}
\widetilde{\Phi_{\omega}^{(1)}}\left(\frac{\pi}{2}\right)=0=\widetilde{\Phi_{\omega}^{(2)}}\left(-\frac{\pi}{2}\right)\,,
\end{align}
which we will refer to as the \emph{Dirichlet type II} condition.
\item $e^{i\alpha_\pm}=1$, \emph{i.e.}, $U=\mathbb{I}$. This self--adjoint boundary condition then takes the form of a Dirichlet boundary condition for the first weighted component of the spinor, namely,
\begin{align}\label{Bc3}
\widetilde{\Phi_{\omega}^{(1)}}\left(\frac{\pi}{2}\right)=0=\widetilde{\Phi_{\omega}^{(1)}}\left(-\frac{\pi}{2}\right)\,.
\end{align}
We will call this boundary condition \emph{Dirichlet type III}.
\item $e^{i\alpha_\pm}=-1$, \emph{i.e.}, $U=-\mathbb{I}$. In this case we have a Dirichlet boundary condition for the second weighted component of the spinor, namely,
\begin{align}\label{Bc4}
\widetilde{\Phi_{\omega}^{(2)}}\left(\frac{\pi}{2}\right)=0=\widetilde{\Phi_{\omega}^{(2)}}\left(-\frac{\pi}{2}\right)\,.
\end{align}
This boundary condition will be referred to as \emph{Dirichlet type IV}.
\end{enumerate}

For the case $M\geq 1/2$ we recall that the unique  self--adjoint extension of the operator $\mathbb{D}$ is its closure $\mathbb{\overline{D}}$. As pointed out at the end of Sec.~\ref{SelfAdjSec}, square--integrable solutions of $\mathbb{D}^{\dagger}\Phi_{\omega}=\omega\Phi_{\omega}$ can be characterized by solutions satisfying the Dirichlet type I boundary condition in Eq.~\eqref{aux2}. This boundary condition is the same as those found in Eq.~\eqref{Bc2} above, and thus, also invariant under the infinitesimal action of $\SLC$.

\subsection{Invariant mode functions}\label{InvSec}

Now that we have found which of the boundary conditions that characterize the admissible self--adjoint extensions of the operator $\mathbb{D}$ are invariant under $\SLC$, we shall find the frequency spectrum and the corresponding mode solutions for each of these boundary conditions. It will be convenient to analyse the massless and massive cases separately.

\subsubsection{Massless field}

We know from the analysis in Sec.~\ref{InvariantBC-Sec} that any boundary condition of the form given by Eq.~\eqref{SABC1} with a diagonal matrix $U$ will result in an invariant self--adjoint extension of $\mathbb{D}$. Let us reparametrize this matrix as $U=\mathrm{diag}(e^{2i\beta_+},e^{2i\beta_-})$, with $0\leq\beta_\pm\leq\pi$, so that the boundary condition for this case is now written as
\begin{align}
\cos\beta_{\pm}\Phi_{\omega}^{(1)}\left(\pm\frac{\pi}{2}\right)=\mp\sin\beta_\pm\Phi_{\omega}^{(2)}\left(\pm\frac{\pi}{2}\right)\,,
\end{align}
which, by Eq.~\eqref{BoundaryValues0}, read
\begin{align}\label{LinearEqs}
\cos\left(\frac{\omega\pi}{2}+\beta_{\pm}\right)\tilde{C}_1\pm\sin\left(\frac{\omega\pi}{2}+\beta_{\pm}\right)\tilde{C}_2&=0\,.
\end{align}
To have non--trivial solutions for $\tilde{C}_1$ and $\tilde{C}_2$, the determinant of the associated linear system of equations should vanish, that is, $\sin\left(\omega\pi+\beta_{+}+\beta_{-}\right)=0$, which means that the frequency $\omega$ is restricted by this condition to be of the form 
\begin{align}\label{BetaFreq}
\omega_{j}=-\frac{1}{\pi}\left(\beta_{+}+\beta_{-}\right)+j\,,\hspace{.5cm}j\in\mathbb{Z}\,.
\end{align}

In order to substitute these values back into Eq.~\eqref{LinearEqs}, we need to treat the cases for which $j$ is even and odd separately. By a direct calculation it can readily be verified that the constants $\tilde{C}_1$ and $\tilde{C}_2$ must then satisfy
\begin{align}
\cos\left(\frac{\beta_{+}-\beta_{-}}{2}\right)\tilde{C}_1+\sin\left(\frac{\beta_{+}-\beta_{-}}{2}\right)\tilde{C}_2&=0\,,\hspace{.5cm}\text{if}\,\,\,\,j=2m\,,\nonumber\\
\sin\left(\frac{\beta_{+}-\beta_{-}}{2}\right)\tilde{C}_1-\cos\left(\frac{\beta_{+}-\beta_{-}}{2}\right)\tilde{C}_2&=0\,,\hspace{.5cm}\text{if}\,\,\,\,j=2m+1\,,
\end{align}
for $m\in\mathbb{Z}$. Substituting back into the solutions, and relabeling the index $m$ to $m+1\in\mathbb{Z}$ for later convenience, we find that the mode functions will be given by
\begin{subequations}\label{Betamodes}
\begin{align}
\Psi_{2m,0}(t,\rho)&=N_{2m}\begin{pmatrix}
\cos\left[\left(2m+1-\beta\right)\rho-B\right]\\[.7em]
-\sin\left[\left(2m+1-\beta\right)\rho-B\right]
\end{pmatrix}e^{-i\left(2m+1-\beta\right)t}\,,\label{Betamodes1}\\
\Psi_{2m+1,0}(t,\rho)&=N_{2m+1}\begin{pmatrix}
\sin\left[\left(2m+2-\beta\right)\rho-B\right]\\[.7em]
\cos\left[\left(2m+2-\beta\right)\rho-B\right]
\end{pmatrix}e^{-i\left(2m+2-\beta\right)t}\,,\label{Betamodes2}
\end{align}
\end{subequations}
where we have defined $\beta:=(\beta_{+}+\beta_{-})/\pi$ and $B:=(\beta_{+}-\beta_{-})/2$. Using the inner product in Eq.~\eqref{innerp2}, it follows that $N_j=\pi^{-1/2}$ for all $j\in\mathbb{Z}$.

\subsubsection{Massive field}\label{MassiveInv}

For the massive spinor field with mass satisfying $0<M<1/2$, we only have four unitary matrices $U$ leading to invariant self--adjoint  extensions as listed at the end of Sec.~\ref{InvariantBC-Sec}. In each of these cases we impose the associated boundary conditions to the general solutions given by Eqs.~\eqref{SolComp1} and~\eqref{SolComp2} and substitute the resulting spatial components into Eq.~\eqref{modedecomp1}. For all the mode functions obtained below, we use the definition of the Jacobi polynomials~\cite{nist} given in terms of the Gaussian hypergeometric function by
\begin{align}\label{Jacobi-def}
P_{n}^{\left(a,b\right)}(x)=\frac{\Gamma(n+a+1)}{n!\Gamma(a+1)}F\left(n+a+b+1,-n;a+1;\frac{1-x}{2}\right)\,,\hspace{.5cm} n\in\mathbb{N}\cup\{0\}.
\end{align}
\begin{enumerate}
\item $U=\mathrm{diag}(-1,1)$: The Dirichlet type I boundary condition in Eq.~\eqref{Bc1} applied to Eq.~\eqref{BoundaryValues} reduces to the requirement $C_2=0$ and $A_1^{(M)}=0$, which restricts the values of $\omega$ to be of the form $\pm\omega_{n}^{I}=\pm(1/2+M+n)$, with $n\in\mathbb{N}\cup\{0\}$. Then the solutions take the form
\begin{subequations}\label{Bc1-Modes}
\begin{align}
\Psi_{n,M}^{I}(t,\rho)&=N_{n,M}^{I}(\cos\rho)^{M}\begin{pmatrix}
(1+\sin\rho)^{\frac{1}{2}}P_{n}^{\left(-\frac{1}{2}+M,\frac{1}{2}+M\right)}(\sin\rho)\\
(1-\sin\rho)^{\frac{1}{2}}P_{n}^{\left(\frac{1}{2}+M,-\frac{1}{2}+M\right)}(\sin\rho)
\end{pmatrix}e^{-i\omega_{n}^{I}t}\,,\label{Bc1-Modes1}\\
\Psi_{-n,M}^{I}(t,\rho)&=N_{n,M}^{I}(\cos\rho)^{M}\begin{pmatrix}
(1+\sin\rho)^{\frac{1}{2}}P_{n}^{\left(-\frac{1}{2}+M,\frac{1}{2}+M\right)}(\sin\rho)\\
-(1-\sin\rho)^{\frac{1}{2}}P_{n}^{\left(\frac{1}{2}+M,-\frac{1}{2}+M\right)}(\sin\rho)
\end{pmatrix}e^{i\omega_{n}^{I}t}\,,\label{Bc1-Modes2}
\end{align}
\end{subequations}
with the normalization constant given by
\begin{align}\label{Bc1-NormConst}
N_{n,M}^{I}=\frac{\sqrt{n!\Gamma(n+2M+1)}}{2^{M+\frac{1}{2}}\Gamma(1/2+M+n)}\,.
\end{align}
\item $U=\mathrm{diag}(1,-1)$: The Dirichlet type II boundary condition given in Eq.~\eqref{Bc2} reduces to $C_1=0$ and $A_1^{(-M)}=0$, which restricts the values of $\omega$ to be of the form $\omega=\omega_n^{I\!I}:=n-M+1/2$, with $n\in\mathbb{N}\cup\{0\}$ for the positive--frequency modes, and $\omega=-\omega_{n}^{I\!I}$ for the negative--frequency modes. Then the solutions are found to be given by
\begin{subequations}\label{Bc2-Modes}
\begin{align}
\Psi_{n,M}^{I\!I}(t,\rho)&=N_{n,M}^{I\!I}(\cos\rho)^{-M}\begin{pmatrix}
(1-\sin\rho)^{\frac{1}{2}}P_{n}^{\left(\frac{1}{2}-M,-\frac{1}{2}-M\right)}(\sin\rho)\\
-(1+\sin\rho)^{\frac{1}{2}}P_{n}^{\left(-\frac{1}{2}-M,\frac{1}{2}-M\right)}(\sin\rho)
\end{pmatrix}e^{-i\omega_n^{I\!I}t}\,,\label{Bc2-Modes1}\\
\Psi_{-n,M}^{I\!I}(t,\rho)&=N_{n,M}^{I\!I}(\cos\rho)^{-M}\begin{pmatrix}
(1-\sin\rho)^{\frac{1}{2}}P_{n}^{\left(\frac{1}{2}-M,-\frac{1}{2}-M\right)}(\sin\rho)\\
(1+\sin\rho)^{\frac{1}{2}}P_{n}^{\left(-\frac{1}{2}-M,\frac{1}{2}-M\right)}(\sin\rho)
\end{pmatrix}e^{i\omega_n^{I\!I}t}\,,\label{Bc2-Modes2}
\end{align}
\end{subequations}
with
\begin{align}\label{Bc2-NormConst}
N_{n,M}^{I\!I}=\frac{\sqrt{n!\Gamma(n-2M+1)}}{2^{\frac{1}{2}-M}\Gamma(1/2-M+n)}\,.
\end{align}
\item $U=\mathbb{I}$: The Dirichlet type III boundary condition for the first component in Eq.~\eqref{Bc3} implies that $C_1=0$ and $B_2^{(M)}=0$. Using the definitions in Eq.~\eqref{Trans-Coeffs} this boundary condition restricts the value of $\omega$ to be either zero or of the form $\omega=\omega_n^{I\!I\!I}:=n$, with $n\in\mathbb{N}$ for the positive--frequency modes and $\omega=-\omega_n^{I\!I\!I}$ for the negative--frequency modes. Substituting this into  Eqs.~\eqref{SolComp1} and~\eqref{SolComp2} we have that the mode solutions $\Psi$ reduce to the forms
\begin{subequations}\label{Bc3-Modes}
\begin{align}
\Psi_{n,M}^{I\!I\!I}(t,\rho)&=N_{n,M}^{I\!I\!I}\left(\frac{1+\sin\rho}{1-\sin\rho}\right)^{\frac{M}{2}}\begin{pmatrix}
\cos\rho\,P_{n-1}^{\left(\frac{1}{2}-M,\frac{1}{2}+M\right)}(\sin\rho)\\
-2P_{n}^{\left(-\frac{1}{2}-M,-\frac{1}{2}+M\right)}(\sin\rho)
\end{pmatrix}e^{-i\omega_n^{I\!I\!I}t}\,,\label{Bc3-Modes1}\\
\Psi_{0,M}^{I\!I\!I}(t,\rho)&=N_{0,M}^{I\!I\!I}\left(\frac{1+\sin\rho}{1-\sin\rho}\right)^{\frac{M}{2}}\begin{pmatrix}
0\\
-2
\end{pmatrix}\,,\label{Bc3-Modes2}\\
\Psi_{-n,M}^{I\!I\!I}(t,\rho)&=N_{n,M}^{I\!I\!I}\left(\frac{1+\sin\rho}{1-\sin\rho}\right)^{\frac{M}{2}}\begin{pmatrix}
\cos\rho\,P_{n-1}^{\left(\frac{1}{2}-M,\frac{1}{2}+M\right)}(\sin\rho)\\
2P_{n}^{\left(-\frac{1}{2}-M,-\frac{1}{2}+M\right)}(\sin\rho)
\end{pmatrix}e^{i\omega_n^{I\!I\!I}t}\,,\label{Bc3-Modes3}
\end{align}
\end{subequations}
where
\begin{align}\label{Bc3-NormConst}
N_{n,M}^{I\!I\!I}=\frac{n!}{2\sqrt{\Gamma(1/2+M+n)\Gamma(1/2-M+n)}}\,.
\end{align}
\item $U=-\mathbb{I}$: The Dirichlet type IV boundary condition in Eq.~\eqref{Bc4} implies that $C_2=0$ and $\omega\,B_2^{(-M)}=0$. Once again, from the definitions in Eq.~\eqref{Trans-Coeffs} these conditions restrict the value of $\omega$ to be either zero or once again of the form $\omega=\omega_n^{I\!I\!I}=n$, with $n\in\mathbb{N}$ for the positive--frequency modes and $\omega=-\omega_{n}^{I\!I\!I}$ for the negative--frequency modes. Thus, the associated mode solutions $\Psi$ reduce to the form
\begin{subequations}\label{Bc4-Modes}
\begin{align}
\Psi_{n,M}^{I\!V}(t,\rho)&=N_{n,M}^{I\!V}\left(\frac{1-\sin\rho}{1+\sin\rho}\right)^{\frac{M}{2}}\begin{pmatrix}
2P_{n}^{\left(-\frac{1}{2}+M,-\frac{1}{2}-M\right)}(\sin\rho)\\
\cos\rho\,P_{n-1}^{\left(\frac{1}{2}+M,\frac{1}{2}-M\right)}(\sin\rho)
\end{pmatrix}e^{-i\omega_n^{I\!I\!I}t}\,,\label{Bc4-Modes1}\\
\Psi_{0,M}^{I\!V}(t,\rho)&=N_{0,M}^{I\!V}\left(\frac{1-\sin\rho}{1+\sin\rho}\right)^{\frac{M}{2}}\begin{pmatrix}
2\\
0
\end{pmatrix}\,,\label{Bc4-Modes2}\\
\Psi_{-n,M}^{I\!V}(t,\rho)&=N_{n,M}^{I\!V}\left(\frac{1-\sin\rho}{1+\sin\rho}\right)^{\frac{M}{2}}\begin{pmatrix}
2P_{n}^{\left(-\frac{1}{2}+M,-\frac{1}{2}-M\right)}(\sin\rho)\\
-\cos\rho\,P_{n-1}^{\left(\frac{1}{2}+M,\frac{1}{2}-M\right)}(\sin\rho)
\end{pmatrix}e^{i\omega_n^{I\!I\!I}t}\,,\label{Bc4-Modes3}
\end{align}
\end{subequations}
with $N_{n,M}^{I\!V}=N_{n,M}^{I\!I\!I}$ as given in Eq.~\eqref{Bc3-NormConst}.
\end{enumerate}

From the remarks at the end of Sec.~\ref{InvariantBC-Sec}, all square--integrable solutions with $M\geq 1/2$ are invariant under $\SLC$ and satisfy the boundary condition in Eq.~\eqref{aux2}. As shown in Sec.~\ref{DiracEqSec}, imposing this boundary condition (or equivalently, requiring square--integrable solutions) restricts the values of the frequencies to be $\omega_{n}^{I}$ and requires the second linearly independent solution to vanish. We substitute these conditions into the general solutions found for $M\geq 1/2$ as follows:
\begin{enumerate}
\item For $M-1/2\notin\mathbb{N}\cup\{0\}$: We substitute $C_2=0$ and $\omega_{n}^{I}=1/2+M+n$, $n\in\mathbb{N}\cup\{0\}$ into the general solutions given by Eq.~\eqref{SolComp1} and~\eqref{SolComp2}. This results in the mode solutions $\Psi_{n,M}^{I}$ in Eq.~\eqref{Bc1-Modes1}. This was indeed expected as both sets satisfy the same boundary condition~\eqref{aux2}, the only difference being the values of the mass $M$ in each in case.
\item For $M=k+1/2$, with $k\in\mathbb{N}$:  We substitute $C_2=0$ and $\omega_{n}^{I}=k+n+1$, $n\in\mathbb{N}\cup\{0\}$ into the general solutions given by Eq.~\eqref{SolCompklog}. This results in the spatial components 
\begin{subequations}
\begin{align}
\Phi_{\omega_{n}^{I}}^{(1)}(\rho)&=C_{1}\sigma(\rho)^{\frac{1}{2}}\left(\mathsf{P}_{k+n+1}^{-k}(\sin\rho)+\mathsf{P}_{k+n}^{-k}(\sin\rho)\right)\,,\\
\Phi_{\omega_{n}^{I}}^{(2)}(\rho)&=C_{1}\sigma(\rho)^{-\frac{1}{2}}\left(\mathsf{P}_{k+n}^{-k}(\sin\rho)-\mathsf{P}_{k+n+1}^{-k}(\sin\rho)\right)\,.
\end{align}
\end{subequations}
Writing the Ferrers functions above in terms of Gaussian hypergeometric functions using Eq.~\eqref{Ferrers} we find that the components above reduce, via Eq.~\eqref{Jacobi-def} to Jacobi polynomials, so that the mode solutions are given by
\begin{align}\label{KSpinor-Mode}
\Psi_{n,k}^{V}(t,\rho)=N_{n,k}^{V}(\cos\rho)^{k+\frac{1}{2}}\begin{pmatrix}
(1+\sin\rho)^{\frac{1}{2}}P_{n}^{(k,k+1)}(\sin\rho)\\
(1-\sin\rho)^{\frac{1}{2}}P_{n}^{(k+1,k)}(\sin\rho)
\end{pmatrix}e^{-i(k+n+1)t}\,,
\end{align}
with
\begin{align}
N_{n,k}^{V}=\frac{\sqrt{n!(2k+n+1)!}}{2^{k+1}(n+k)!}\,.
\end{align}
From these expressions it is clear that the mode solutions $\Psi_{n,k}^{V}$ are of the same form as $\Psi_{n,M}^{I}$ in Eq.~\eqref{Bc1-Modes1} with $M=k+1/2$.
\item For $M=1/2$: Substituting $C_2=0$ and $\omega_{n}^{I}=m$, where $m\in\mathbb{N}$ into Eq.~\eqref{SolCompklog} with $k=0$, we find the spatial components of the spinor solutions as
\begin{subequations}
\begin{align}
\Phi_{\omega_{m}^{I}}^{(1)}&=C_1\sigma(\rho)^{\frac{1}{2}}\left(\mathsf{P}_{m+1}(\sin\rho)+\mathsf{P}_{m}(\sin\rho)\right)\,,\\
\Phi_{\omega_{m}^{I}}^{(2)}&=C_1\sigma(\rho)^{-\frac{1}{2}}\left(\mathsf{P}_{m}(\sin\rho)-\mathsf{P}_{m+1}(\sin\rho)\right)\,.
\end{align}
\end{subequations}
To match the functional form of these component functions to the previous cases we use the fact that Legendre polynomials are related to Jacobi polynomials by $\mathsf{P}_m(x)=P_{m}^{(0,0)}(x)$. Applying recursion relations for the combinations above it can readily be verified that the resulting mode solutions reduce to
\begin{align}\label{HalkfMass-Mode}
\Psi_{n,1/2}^{V\!I}(t,\rho)=N_{n}^{V\!I}(\cos\rho)^{\frac{1}{2}}\begin{pmatrix}
(1+\sin\rho)^{\frac{1}{2}}P_{n}^{(0,1)}(\sin\rho)\\
(1-\sin\rho)^{\frac{1}{2}}P_{n}^{(1,0)}(\sin\rho)
\end{pmatrix}e^{-i(n+1)t}\,,
\end{align}
where $n\in\mathbb{N}\cup\{0\}$, and $N_{n}^{V\!I}=\sqrt{n+1/2}$. From these expressions it follows that these modes are of the form of $\Psi_{n,M}^{I}$ (and thus, of $\Psi_{n,k}^{V}$) with $M=1/2$ ($k=0$, respectively).
\end{enumerate}

Thus, the mode solutions found for all possible values of $M\geq 1/2$ reduce to the form of the spinors $\Psi_{n,M}^{I}$ as given by Eq.~\eqref{Bc1-Modes1}.

\subsection{Mode functions leading to invariant positive--frequency subspaces}\label{UIR-Sec}

We will now determine which of the solution spaces that result from the $\SLC$--invariant self-adjoint boundary conditions found in the previous section split into invariant positive-- and negative--frequency subspaces and thus, lead to an invariant vacuum state via the Fock space construction outlined in Appendix~\ref{App2}. It is clear that all the sets of mode solutions listed in Sec.~\ref{InvSec} form a unitary representation of $\SLC$ with respect to the inner product in Eq.~\eqref{innerp3}, as from Eq.~\eqref{Ladderaction} we have $L_{\pm}^{\dagger}=-L_{\mp}$, and $\left|\!\left|\Psi_{\omega,M}\right|\!\right|^{2}_{2}>0$ for all values of $M$ and all $\omega$ in the frequency spectrum. In order to determine if any of these representations admits a splitting into invariant positive-- and negative--frequency subspaces, we will use the action of the operators $L_{\pm}$ from Eqs.~\eqref{Ladderaction} and~\eqref{Transfcomp-noDer} on each of the sets of mode solutions to see if a particular mode is annihilated by any of these operators and thus, defines a highest or lowest weight vector of an invariant subspace. Once again it will be more convenient to treat the massless and massive cases separately.

{\bf Massless spinor $M=0$:}
We consider the modes in Eq.~\eqref{Betamodes} collectively written as  $\Psi_{j,0}$ for $j\in\mathbb{Z}$. By applying the operators $L_{\pm}$ on both $\Psi_{2n,0}$ and $\Psi_{2n+1,0}$, it can readily be verified that
\begin{align}\label{Trans-Beta-Un}
(L_{\pm}\Psi_{j,0})(t,\rho)=i(-1)^{j+1}\left(\frac{1}{2}\pm\omega_{j}\right)\Psi_{j\pm 1,0}(t,\rho)\,.
\end{align}
We recall that the frequencies are given by $\omega_{j}=j+1-\beta$, with $\beta=(\beta_{+}+\beta_{-})/\pi$ so that $0\leq\beta<2$. This implies that the right--hand side of the expression above vanishes only for $\beta=1/2$ or $\beta=3/2$. This implies that for any other possible value of $\beta$ and $j\in\mathbb{Z}$, the representation is irreducible. In fact, from the classification at the end of Sec.~\ref{AdSsection}, we identify these irreducible representations as unitary principal series of the form $\mathscr{P}_{0}^{\mu}$, with the values of $\mu$ given by 
\begin{align}\label{MuParam}
\mu=\left\{\begin{matrix}
-\beta, & 0\leq\beta<\frac{1}{2}\,\\[.5em]
1-\beta, & \frac{1}{2}<\beta<\frac{3}{2}\,,\\[.5em]
2-\beta, & \frac{3}{2}<\beta<2\,.
\end{matrix}\right.
\end{align}

Now we turn to the specific cases in which the resulting representations are reducible. When $\beta=1/2$, we see from Eq.~\eqref{Trans-Beta-Un} that $L_{+}\Psi_{-1,0}$ and $L_{-}\Psi_{0,0}$, with frequencies $\omega_{-1}=-1/2$ and $\omega_{0}=1/2$ respectively, vanish. Therefore, the representation splits into the two invariant subspaces spanned by the positive--frequency modes $\left\{\Psi_{n,0}\right\}_{n\in\mathbb{N}\cup\{0\}}$, and the negative--frequency modes $\left\{\Psi_{-n,0}\right\}_{n\in\mathbb{N}}$, respectively. The explicit form of the mode solutions is obtained by writing Eq.~\eqref{Betamodes} in terms of one of the parameters, say $\beta_{+}$, so that $\beta_{-}=\pi/2-\beta_{+}\geq 0$ and $B=\beta_{+}-\pi/4$. In this way, for $\beta_{+}\in[0,\pi/2)$ and $n\in\mathbb{N}\cup\{0\}$, the invariant positive--frequency subspace is spanned by the modes
\begin{subequations}\label{BetaPlusLower-Modes}
\begin{align}
\Psi_{2n,0}^{\beta_{+}}(t,\rho)&=\frac{1}{\sqrt{\pi}}R\left(\frac{\pi}{4}-\beta_{+}\right)\begin{pmatrix}
\cos\left(2n+\frac{1}{2}\right)\rho\\[.5em]
-\sin\left(2n+\frac{1}{2}\right)\rho
\end{pmatrix}e^{-i\left(2n+\frac{1}{2}\right)t}\,,\label{BetaPlusLower-Modes1}\\
\Psi_{2n+1,0}^{\beta_{+}}(t,\rho)&=\frac{1}{\sqrt{\pi}}R\left(\frac{\pi}{4}-\beta_{+}\right)\begin{pmatrix}
\sin\left(2n+1+\frac{1}{2}\right)\rho\\[.5em]
\cos\left(2n+1+\frac{1}{2}\right)\rho
\end{pmatrix}e^{-i\left(2n+1+\frac{1}{2}\right)t}\,,\label{BetaPlusLower-Modes2}
\end{align}
\end{subequations}
where we have defined $R(\theta)$ as the $2\times 2$ rotation matrix parametrized by the angle $\theta$. The negative--frequency subspace is obtained by taking $-n\in\mathbb{N}$. 

Similarly, when $\beta=3/2$, the transformed modes $L_{+}\Psi_{0,0}$ and $L_{-}\Psi_{1,0}$, with frequencies $\omega_0=-1/2$ and $\omega_{1}=1/2$ respectively, are the only ones vanishing, and thus, the representation once again splits into the two invariant subspaces spanned by $\left\{\Psi_{n,0}\right\}_{n\in\mathbb{N}}$ and $\left\{\Psi_{-n,0}\right\}_{n\in\mathbb{N}\cup\{0\}}$. The explicit form of the mode solutions is once again obtained by writing Eq.~\eqref{Betamodes} in terms of $\beta_{+}$, so that $\beta_{-}=3\pi/2-\beta_{+}<\pi$ and $B=\beta_{+}-3\pi/4$. Also, by shifting the labels of the spinors so that the lowest positive--frequency mode is $\Psi_{0,0}$, we have that, for $\beta_{+}>\pi/2$ and $n\in\mathbb{N}\cup\{0\}$, the positive--frequency subspace is spanned by
\begin{subequations}\label{BetaPlusUpper-Modes}
\begin{align}
\Psi_{2n,0}^{\beta_{+}}(t,\rho)&=\frac{1}{\sqrt{\pi}}R\left(\frac{3\pi}{4}-\beta_{+}\right)\begin{pmatrix}
\sin\left(2n+\frac{1}{2}\right)\rho\\[.5em]
\cos\left(2n+\frac{1}{2}\right)\rho
\end{pmatrix}e^{-i\left(2n+\frac{1}{2}\right)t}\,,\label{BetaPlusUpper-Modes1}\\
\Psi_{2n+1,0}^{\beta_{+}}(t,\rho)&=\frac{1}{\sqrt{\pi}}R\left(\frac{3\pi}{4}-\beta_{+}\right)\begin{pmatrix}
\cos\left(2n+1+\frac{1}{2}\right)\rho\\[.5em]
-\sin\left(2n+1+\frac{1}{2}\right)\rho
\end{pmatrix}e^{-i\left(2n+1+\frac{1}{2}\right)t}\,,\label{BetaPlusUpper-Modes2}
\end{align}
\end{subequations}
and the negative--frequency modes are obtained by considering instead $-n\in\mathbb{N}$. However, by writing $R(3\pi/4-\beta_{+})=R(\pi/4-\beta_{+})R(\pi/2)$ above, and noting that the matrix $R(\pi/2)$ maps a two component spinor $(a,b)^{T}$ to $(b,-a)^{T}$, it follows that the modes in Eq.~\eqref{BetaPlusUpper-Modes} actually reduce to the same form of the modes in Eq.~\eqref{BetaPlusLower-Modes} (up to a minus sign for the odd modes). Therefore, regardless of the value of $\beta_{+}\in[0,\pi)$, the invariant subspaces the representation splits into are given by the linear span of the modes appearing in Eq.~\eqref{BetaPlusLower-Modes}. 

Before identifying the resulting subspaces with the known UIR's, we will write the mode solutions in terms of Jacobi polynomials so that we can match the functional form of the spinors with $M\neq 0$ found in Sec.~\ref{MassiveInv}. If we use the fact that 
\begin{subequations}\label{Trig-Jacobi}
\begin{align}
\cos\left(n+\frac{1}{2}\right)\rho=&C_{n}\sqrt{\frac{2}{\pi}}\left((-1)^{n}(1+\sin\rho)^{\frac{1}{2}}P_{n}^{\left(-\frac{1}{2},\frac{1}{2}\right)}(\sin\rho)\right.\nonumber\\
&\left.+(1-\sin\rho)^{\frac{1}{2}}P_{n}^{\left(\frac{1}{2},-\frac{1}{2}\right)}(\sin\rho)\right)\,,\\
\sin\left(n+\frac{1}{2}\right)\rho=&C_{n}\sqrt{\frac{2}{\pi}}\left((1+\sin\rho)^{\frac{1}{2}}P_{n}^{\left(-\frac{1}{2},\frac{1}{2}\right)}(\sin\rho)\right.\nonumber\\
&\left.+(-1)^{n}(1-\sin\rho)^{\frac{1}{2}}P_{n}^{\left(\frac{1}{2},-\frac{1}{2}\right)}(\sin\rho)\right)\,,
\end{align}
\end{subequations}
where we have defined the constants
\begin{align}
C_{n}:=\frac{(-1)^{\lfloor \frac{n}{2}\rfloor}n!}{\sqrt{2}\Gamma\left(n+\frac{1}{2}\right)}\,,
\end{align}
then a straightforward calculation shows that the modes in Eq.~\eqref{BetaPlusLower-Modes} are collectively written as
\begin{subequations}\label{Pos-Freq-BetaModes}
\begin{align}
\Psi_{n,0}^{\beta_{+}}(t,\rho)&=C_n R\left(\frac{\pi}{2}-\beta_{+}\right)\begin{pmatrix}
(1+\sin\rho)^{\frac{1}{2}}P_{n}^{\left(-\frac{1}{2},\frac{1}{2}\right)}(\sin\rho)\\[.5em]
(1-\sin\rho)^{\frac{1}{2}}P_{n}^{\left(\frac{1}{2},-\frac{1}{2}\right)}(\sin\rho)
\end{pmatrix}e^{-i\left(\frac{1}{2}+n\right)t}\,,\label{Pos-Freq-BetaModes1}\\
\Psi_{-n,0}^{\beta_{+}}(t,\rho)&=C_n R\left(\frac{\pi}{2}-\beta_{+}\right)\begin{pmatrix}
(1+\sin\rho)^{\frac{1}{2}}P_{n}^{\left(-\frac{1}{2},\frac{1}{2}\right)}(\sin\rho)\\[.5em]
-(1-\sin\rho)^{\frac{1}{2}}P_{n}^{\left(\frac{1}{2},-\frac{1}{2}\right)}(\sin\rho)
\end{pmatrix}e^{i\left(\frac{1}{2}+n\right)t}\,,\label{Pos-Freq-BetaModes2}
\end{align}
\end{subequations}
for all $\beta_{+}\in[0,\pi)$ and $n\in\mathbb{N}\cup\{0\}$. We note that the matrix $R(\theta)$ can be written as $R(\theta)=\exp(-2i\theta\Sigma^{01})$ by means of Eq.~\eqref{sigmamatrix}. Now, it is a well known fact that the massless Dirac equation has a global internal chiral symmetry.~\cite{collas}. In  a two--dimensional spacetime, chirality corresponds to the spinor components of the solutions to Eq.~\eqref{DiracEq1} with $M=0$ being left-- or right--moving plane waves. In terms of the gamma matrix representation in Eq.~\eqref{GammaMatrices}, the chiral transformation is given by $\Psi\mapsto\exp(2i\theta\Sigma^{01})\Psi$ and two massless Dirac spinors differing by a chiral transformation are taken to be equivalent under this symmetry. Then, the action of the rotation matrix $R(\pi/2-\beta)$ in Eq.~\eqref{Pos-Freq-BetaModes} is in fact a chiral transformation on the modes $\Psi_{\pm n,0}^{\beta_{+}}$. Therefore, the sets of mode functions with different values of $\beta_{+}$ define a unique representation up to chiral equivalence. Referring to the classification of UIR's in Sec.~\ref{AdSsection}, we can now directly identify the linear span of the positive--frequency mode solutions in Eq.~\eqref{Pos-Freq-BetaModes1} with the positive mock--discrete series representation $\mathscr{D}_{1/2}^{+}$. Similarly, the negative--frequency subspace spanned by the modes in Eq.~\eqref{Pos-Freq-BetaModes2} is identified with the negative mock--discrete series $\mathscr{D}_{1/2}^{-}$. The fact that the unitary representation spanned by both positive-- and negative--frequency subspaces splits into the two invariant subspaces is consistent with the representation theory of $\SLC$. The reducible representation spanned by both positive-- and negative--frequency modes corresponds to the unitary principal series $\mathscr{P}_{0}^{1/2}$, which is known to have the decomposition into irreducible subspaces $\mathscr{P}_{0}^{1/2}\simeq \mathscr{D}^{+}_{1/2}\oplus\mathscr{D}^{-}_{1/2}$.~\cite{knapp,puk}

We also note that the invariant sets of mode functions corresponding to the Dirichlet types I--IV boundary conditions in Eqs.~\eqref{Bc1-Modes}--\eqref{Bc4-Modes} reduce to certain massless sets of modes in the limit  $M\to 0$. The Dirichlet type I and type II modes with $M=0$ reduce to Eq.~\eqref{Pos-Freq-BetaModes} with $\beta_{+}=\pi/2$ and $\beta_{+}=0$, respectively. To see the correspondence with the Dirichlet type III and type IV modes, we consider the invariant massless mode functions in Eq.~\eqref{Betamodes} forming the principal series $\mathscr{P}_{0}^{\mu}$, with $\mu=0$. From Eq.~\eqref{MuParam} this restricts the values of $\beta_\pm$ to satisfy $\beta=1$. Thus, by setting $\beta_{-}=\pi-\beta_{+}$ in Eq.~\eqref{Betamodes}, the invariant massless modes reduce to
\begin{subequations}\label{BetamodesPrin}
\begin{align}
\Psi_{2m,0}(t,\rho)&=\frac{1}{\sqrt{\pi}}R(-\beta_{+})\begin{pmatrix}
-\sin\, 2m\rho\\[.7em]
-\cos\, 2m\rho
\end{pmatrix}e^{-2imt}\,,\label{BetamodesPrin1}\\
\Psi_{2m+1,0}(t,\rho)&=\frac{1}{\sqrt{\pi}}R(-\beta_{+})
\begin{pmatrix}
\cos(2m+1)\rho\\[.7em]
-\sin(2m+1)\rho
\end{pmatrix}e^{-i\left(2m+1\right)t}\,,\label{BetamodesPrin2}
\end{align}
\end{subequations}
where $R(\beta_{+})$ is a rotation matrix by the angle $\beta_{+}$. We now write these modes in a form more readily recognizable as the massless limit of the massive cases. Consider the identities $\cos \,n\theta=T_{n}(\cos\theta)$, and $\sin\,n\theta=\sin\theta U_{n-1}(\cos\theta)$, where $n\in\mathbb{N}\cup\{0\}$ and $T_{n}$, $U_n$ are Chebyshev polynomials~\cite{nist} of the first and second kind, respectively, and the relations
\begin{align}\label{Chebyshev-Id}
T_n(x)=\frac{n!\sqrt{\pi}}{\Gamma(n+1/2)}P_{n}^{(-1/2,-1/2)}(x)\,,\hspace{.5cm}U_n(x)=\frac{(n+1)!\sqrt{\pi}}{2\Gamma(n+3/2)}P_{n}^{(1/2,1/2)}(x)\,.
\end{align}
Using these identities, one finds that Eq.~\eqref{BetamodesPrin} is written, for $n\in\mathbb{N}$, as
\begin{subequations}\label{FinalBetamodes}
\begin{align}
\Psi_{n,0}(t,\rho)&=(-1)^{n}N_{n,0}^{III}R(-\beta_{+})\begin{pmatrix}
\cos\rho\,P_{n-1}^{(1/2,1/2)}(\sin\rho)\\[.7em]
-2P_{n}^{(-1/2,-1/2)}(\sin\rho)
\end{pmatrix}e^{-int}\,,\\
\Psi_{0,0}(t,\rho)&=N_{0,0}^{III}R(-\beta_{+})\begin{pmatrix}
0\\[.7em]
-2
\end{pmatrix}\,,\\
\Psi_{-n,0}(t,\rho)&=(-1)^{n}N_{n,0}^{III}R(-\beta_{+})\begin{pmatrix}
-\cos\rho\,P_{n-1}^{(1/2,1/2)}(\sin\rho)\\[.7em]
-2P_{n}^{(-1/2,-1/2)}(\sin\rho)
\end{pmatrix}e^{int}\,.
\end{align}
\end{subequations}
From these expressions it follows that if $\beta_{+}=0$, then the modes above are the Dirichlet type III mode solutions in Eq.~\eqref{Bc3-Modes} with $M=0$, and if $\beta_{+}=\pi/2$, then we obtain the Dirichlet type IV mode solutions in Eq.~\eqref{Bc4-Modes} with $M=0$. Similarly to the previous situation, all other values of $\beta_{+}$ are equivalent to either of these two mode solutions up to a chiral transformation.

{\bf Massive spinor:} Now we analyze the massive mode solutions that resulted from imposing the self--adjoint boundary conditions which are invariant under $\SLC$. We will determine which of the sets of mode solutions appearing in Eqs.~\eqref{Bc1-Modes}--\eqref{Bc4-Modes} span a solution space with an invariant positive-- or negative--frequency subspace by applying the ladder operators $L_{\pm}$ on the lowest positive-- and highest negative--frequency modes and see for which cases these are annihilated.

We first consider the two sets of mode solutions $\Psi_{n,M}^{I}$ for all $M>0$ and $\Psi_{n,M}^{I\!I}$ for $0<M<1/2$, given by Eqs.~\eqref{Bc1-Modes} and~\eqref{Bc2-Modes}. Using Eq.~\eqref{Ladderaction} and standard recurrence relations for the Jacobi polynomials~\cite{nist} we find that the set of Dirichlet type I  modes transform under the ladder operators $L_\pm$ as
\begin{subequations}\label{LadderModeActionI}
\begin{align}
L\pm\Psi_{n,M}^{I}=-i\sqrt{(n+1/2\pm 1/2)(n+2M+1/2\pm 1/2)}\Psi_{n\pm 1,M}^{I}\,,\\
L\pm\Psi_{-n,M}^{I}=i\sqrt{(n+1/2\mp 1/2)(n+2M+1/2\mp 1/2)}\Psi_{-n\pm 1,M}^{I}\,,
\end{align}
\end{subequations}
for all $n\geq 0$ (here we use the notation $\Psi_{-0,M}^{I}$ to denote the highest negative--frequency mode). Similarly, the Dirichlet type II mode solutions transform as in Eq.~\eqref{LadderModeActionI} with $M$ replaced by $-M$.  From these expressions we see that $L_{-}\Psi_{0,M}=0=L_{+}\Psi_{-0,M}$, and thus, both the positive-- and negative--frequency subspaces are invariant for these two sets of mode solutions. Furthermore, using the classification of UIR's at the end of Sec.~\ref{AdSsection}, we can identify these subspaces with the discrete series representations: The mode solutions $\Psi_{\pm n,M}^{I}$ form the representation $\mathscr{D}^{\pm}_{1/2+M}$, while the mode solutions $\Psi_{\pm n,M}^{I\!I}$ form the representation $\mathscr{D}^{\pm}_{1/2-M}$.

On the other hand, Dirichlet type III and type IV  mode solutions in Eqs.~\eqref{Bc3-Modes} and~\eqref{Bc4-Modes}, respectively, form an irreducible representation and thus, do not split into invariant positive-- and negative--frequency subspaces. This can be seen as follows: Using Eq.~\eqref{Ladderaction} we find that both sets of mode solutions, $\Psi_{n}^{I\!I\!I}$ and $\Psi_{n}^{I\!V}$, transform under the action of the ladder operators $L_{\pm}$ as
\begin{subequations}\label{Mode-LaddTrans}
\begin{align}
L_{\pm}\Psi_{n}&=-i\sqrt{(n+M\pm 1/2)(n-M\pm 1/2)}\Psi_{n\pm 1}\,,\\
L_{\pm}\Psi_{-n}&=i\sqrt{(n+M\mp 1/2)(n-M\mp 1/2)}\Psi_{-n\pm 1}\,,
\end{align}
\end{subequations}
for all $n\geq 1$, and 
\begin{align}\label{Mode-LaddTrans0}
L_\pm \Psi_{0}=\mp i\sqrt{1/4-M^2}\Psi_{\pm 1}\,.
\end{align}
Because these solutions are valid only for $0<M<1/2$, the right--hand side of Eqs.~\eqref{Mode-LaddTrans} and~\eqref{Mode-LaddTrans0} are never zero for any $n\in\mathbb{N}\cup\{0\}$, and thus, all consecutive modes appearing in Eqs.~\eqref{Bc3-Modes} and~\eqref{Bc4-Modes} can be reached by applying ladder operators $L_{\pm}$, so the associated spaces spanned by these modes are invariant under the action of $\SLC$. In fact, using the classification of UIRs, we can identify both of  these solution spaces with the complementary series representations $\mathscr{C}^{0}_{1/2+M}$.

We have found that only the mode solutions stemming from the invariant self--adjoint boundary conditions that span a solution space with invariant positive-- and negative--frequency subspaces are those coming from Dirichlet type I boundary condition given by Eq.~\eqref{Bc1-Modes} and from the Dirichlet type II boundary condition given by Eq.~\eqref{Bc2-Modes}, which correspond to the self--adjoint extensions of the operator $\mathbb{D}$ labelled by the matrices $U=\mathrm{diag}(\mp 1,\pm 1)$, respectively. 

It is also worth noting that the massless and massive mode solutions that result from imposing the Dirichlet type I--IV boundary conditions are invariant under charge conjugation, $\Psi\mapsto \Psi^{c}=C(\gamma^{0})^{T}\overline{\Psi}$ with $C=-2\Sigma^{01}$. This fact follows immediately by noting that all of the negative--frequency modes from these sets of mode solutions satisfy $\Psi_{-n}=-\gamma^{1}\overline{\Psi_{n}}$ for all $n\geq 0$. From Eqs.~\eqref{GammaMatrices} and~\eqref{sigmamatrix}, we have $C(\gamma^{0})^{T}=-\gamma^{1}$ and thus, all Dirichlet type I--IV modes satisfy $\Psi_{n}^{c}=\Psi_{-n}$. Furthermore, the only self--adjoint boundary conditions that are invariant under both, charge conjugation and $\SLC$--transformations, are those corresponding to the four unitary matrices $U=\mathrm{diag}(\pm 1,\mp 1)$ and $U=\pm \mathbb{I}$. A simple calculation shows that if $\Phi_{\omega}$ satisfies the general self--adjoint boundary condition of Eq.~\eqref{SABC1}, then the charge conjugate $\Phi_{\omega}^{c}=-\gamma^{1}\Phi_{\omega}$ satisfies the same boundary condition if and only if the matrix $U$ satisfies $\overline{U}=U$. We can then determine which of these charge conjugation--invariant boundary conditions are also invariant under $\SLC$--transformations using the analysis of Sec.~\ref{InvariantBC-Sec} but assuming $\overline{U}=U$. We find that if $M=0$ the matrix $U$ must also be diagonal and the only unitary matrices satisfying both of these requirements are $U=\mathrm{diag}(\pm 1,\mp 1)$ and $U=\pm \mathbb{I}$. If $M\neq 0$, we find the same four unitary matrices. 

\subsection{Isometry-invariant theories with no invariant positive-frequency subspaces}\label{NonInv-Sec}

In Sec.~\ref{UIR-Sec} we have found that only certain $\SLC$--invariant self--adjoint boundary conditions result in invariant positive--frequency subspaces, and thus, in an invariant vacuum state once the Fock space construction, as outlined in Appendix~\ref{App2}, is performed. We also noted that the rest of the $\SLC$--invariant self--adjoint boundary conditions result in unitary representations that do not split into positive-- or negative--frequency subspaces and thus, no invariant vacuum state can be found. Instead, because the inner product~\eqref{innerp1} is $\SLC$--invariant for any of these boundary conditions, the ladder operators $L_{\pm}$ acting on the quantum field are  Bogoliubov transformations that mix the creation and annihilation operators. This implies that for these theories there must be UIRs of $\SLC$ to which the associated vacuum states belong to. In this section we find these representations.

We will start by considering the massless modes in Eq.~\eqref{Betamodes}. Without loss of generality, we will choose the parameters $\beta_{\pm}$ of the unitary matrix $U$ such that $\omega_{0}=\mu>0$ is the lowest positive frequency. The analysis for the other representations labelled by $\mu$ in Eq.~\eqref{MuParam} can be carried over by appropriately relabeling the frequency index. The case $\mu=0$ will be analyzed separately.

We recall that the frequency spectrum is given by $\omega_{j}=j+\mu$, $j\in\mathbb{Z}$, and for the sake of simplicity, we will denote the associated mode solutions by $\Psi_{j}$ instead of $\Psi_{j,0}$.  The negative--frequency modes are given by $\Psi_{-j}$ for $j>0$ and, thus, the quantum field $\Psi$ is expanded in terms of the complete set of mode solutions as
\begin{align}\label{Quantum-Field-Exp}
\Psi=\sum_{j\geq 0}\left(a_{j}\Psi_{j}+b_{j}^{\dagger}\Psi_{-j-1}\right)\,,
\end{align}
where the operators $a_{j},b_{k}$ satisfy
\begin{align}\label{CACRs}
\left\{a_{j},a_{k}^{\dagger}\right\}=\delta_{jk}\mathbb{I}=\left\{b_{j},b_{k}^{\dagger}\right\}\,,
\end{align}
and all other anticommutators vanish. Using the action of the ladder operators on the modes $\Psi_{j}$ given by Eq.~\eqref{Trans-Beta-Un}, we find that 
\begin{subequations}\label{QF-Ladder-Ops}
\begin{align}
L_{+}\Psi=&i\sum_{j\geq 0}(-1)^{j+1}\left(\left(\omega_{j}+\frac{1}{2}\right)a_{j}\Psi_{j+1}+\left(\omega_{-j-1}-\frac{1}{2}\right)b_{j+1}^{\dagger}\Psi_{-j-1}\right)\nonumber\\
&+i\left(\mu-\frac{1}{2}\right)b_{0}^{\dagger}\Psi_{0}\,,\\
L_{-}\Psi=&i\sum_{j\geq 0}(-1)^{j+1}\left(\left(\omega_{j}+\frac{1}{2}\right)a_{j+1}\Psi_{j}+\left(\omega_{-j-1}-\frac{1}{2}\right)b_{j}^{\dagger}\Psi_{-j-2}\right)\nonumber\\
&+i\left(\mu-\frac{1}{2}\right)a_{0}\Psi_{-1}\,.
\end{align}
\end{subequations}

Using the inner product in Eq.~\eqref{innerp1} in terms of the field $\Psi=(\cos\rho)^{-1/2}\psi$, for which the mode solutions $\Psi_{j}$ satisfy $\inpr{\Psi_j}{\Psi_k}_{D}=\delta_{jk}$, we define the conserved quantum charges for the symmetry generated by $L_\pm$ by
\begin{align}\label{QuantumCharges}
\hat{L}_{\pm}:=\inpr{\Psi}{L_{\pm}\Psi}_{D}\,,
\end{align}
and from Eq.~\eqref{QF-Ladder-Ops}, we find that these can be written in terms of the annihilation and creation operators as
\begin{subequations}\label{Fock-Ladder-Ops}
\begin{align}
\hat{L}_{+}&=i\sum_{j\geq 0}(-1)^{j+1}\!\!\left(\left(\omega_{j}+\frac{1}{2}\right)a_{j+1}^{\dagger}a_{j}+\left(\frac{1}{2}-\omega_{-j-1}\right)b_{j+1}^{\dagger}b_{j}\right)+i\left(\mu-\frac{1}{2}\right)a_{0}^{\dagger}b_{0}^{\dagger}\,,\\
\hat{L}_{-}&=i\sum_{j\geq 0}(-1)^{j+1}\!\!\left(\left(\omega_{j}+\frac{1}{2}\right)a_{j}^{\dagger}a_{j+1}+\left(\frac{1}{2}-\omega_{-j-1}\right)b_{j}^{\dagger}b_{j+1}\right)-i\left(\mu-\frac{1}{2}\right)a_{0}b_{0}\,.\label{Fock-Ladder-Ops2}
\end{align}
\end{subequations}
Next we calculate the commutator between these charges. 
With the anticommutation relations in Eq.~\eqref{CACRs}, it can readily be verified that 
\begin{align}\label{Fock-Comm1}
\left[\hat{L}_{+},\hat{L}_{-}\right]=2\sum_{j\geq 0}\left(\omega_{j}a^{\dagger}_{j}a_{j}-\omega_{-j-1}b_{j}^{\dagger}b_{j}\right)+\left(\mu-\frac{1}{2}\right)^2\mathbb{I}\,.
\end{align}
We then define the operator 
\begin{align}
\hat{L}_{0}:=\sum_{j\geq 0}\left(\omega_{j}a^{\dagger}_{j}a_{j}-\omega_{-j-1}b_{j}^{\dagger}b_{j}\right)+2\lambda\mathbb{I}\,,
\end{align} 
with $\lambda=(\mu-1/2)^{2}/2$. If we then compare Eq.~\eqref{Fock-Comm1} with Eq.~\eqref{Comm-LadderOps}, we can identify $\hat{L}_{0}$ with the time--translation charge induced from $\mathcal{L}_{0}$. Following the construction in Appendix~\ref{App2}, the vacuum state $\ket{0}$ is defined by the requirement that for all $j\geq 0$, $a_{j}\ket{0}=0=b_{j}\ket{0}$, and we see that
\begin{align}
\hat{L}_{0}\ket{0}=\lambda\ket{0}\,,\hspace{.5cm}\hat{L}_{-}\ket{0}=0\,,
\end{align}
the latter resulting directly from Eq.~\eqref{Fock-Ladder-Ops2}. The Fock space is thus a weight--module with lowest weight $\lambda$ and lowest weight--vector $\ket{0}$. From the classification of UIRs at the end of Sec.~\ref{AdSsection} we see that this representation is isomorphic to the discrete series $\mathscr{D}^{+}_{\lambda}$.

We now turn to the analysis of the massless modes with $\mu=0$ and the massive modes with $0< M<1/2$ satisfying the self--adjoint boundary condition with $U=\mathbb{I}$, given by Eq. ~\eqref{Bc3-Modes}. We recall the fact that the massless modes with $\mu=0$ can be written as in Eq.~\eqref{FinalBetamodes} and thus, are equivalent to the Dirichlet type III modes with $M=0$ up to a chiral transformation, therefore, this analysis includes the case for which $M=0$ and $\mu=0$. Furthermore, the massive Dirichlet type IV modes in Eq.~\eqref{Bc4-Modes} are related to the Dirichlet type III modes by $\Psi^{I\!V}_{n}=(-1)^{n}\mathbb{P}\Psi_{n}^{I\!I\!I}$, where $\mathbb{P}\Psi(t,\rho)=i\gamma^{0}\Psi(t,-\rho)$ is the parity transformation acting on the spinor $\Psi$. Noting that $\mathbb{P}L_{\pm}\mathbb{P}=-L_{\pm}$, it follows that the ladder operators take the same form for the Dirichlet type IV modes as for the Dirichlet type III modes. Thus, without loss of generality, we will consider the mode solutions $\Psi_{\pm n}^{I\!I\!I}$ with $0\leq M<1/2$ to include all remaining cases. For these theories, the quantum field is expanded as
\begin{align}\label{Quantum-ModeExp}
\Psi^{I\!I\!I}=\sum_{n=1}^{\infty}\left(a_{n}\Psi_{n}^{I\!I\!I}+b_{n}^{\dagger}\Psi_{-n}^{I\!I\!I}\right)+a_{0}\Psi_{0}^{I\!I\!I}\,,
\end{align}
with the annihilation and creation operators satisfying Eq.~\eqref{CACRs}. 
Using the transformation in Eq.~\eqref{Mode-LaddTrans} we find that the action of the ladder operators on the quantum field is given by
\begin{subequations}
\begin{align}
L_{+}\Psi^{I\!I\!I}=&-i\sum_{n=1}^{\infty}C_{n}\left(a_{n}\Psi_{n+1}^{I\!I\!I}-b_{n+1}^{\dagger}\Psi_{-n}^{I\!I\!I}\right)+iC_{0}\left(b_{1}^{\dagger}\Psi^{I\!I\!I}_{0}-a_{0}\Psi_{1}^{I\!I\!I}\right)\,,\\
L_{-}\Psi^{I\!I\!I}=&-i\sum_{n=1}^{\infty}C_{n}\left(a_{n+1}\Psi_{n}^{I\!I\!I}-b_{n}^{\dagger}\Psi_{-n-1}^{I\!I\!I}\right)+iC_{0}\left(a_{0}\Psi^{I\!I\!I}_{-1}-a_{1}\Psi_{0}^{I\!I\!I}\right)\,,
\end{align}
\end{subequations}
where we have defined the constants $C_{n}=\sqrt{(n+M+1/2)(n-M+1/2)}$.

Once again, we define the conserved quantum charges $\hat{L}_{\pm}$ by Eq.~\eqref{QuantumCharges}, and considering the fact that the mode solutions $\Psi_{n}^{I\!I\!I}$ are also orthonormal with respect to the inner product \eqref{innerp1}, we find the quantum operators $\hat{L}_{\pm}$ to be given by
\begin{subequations}
\begin{align}
\hat{L}_{+}&=-i\sum_{n=1}^{\infty}C_{n}\left(a_{n+1}^{\dagger}a_{n}+b_{n+1}^{\dagger}b_{n}\right)+iC_{0}\left(a_{0}^{\dagger}b_{1}^{\dagger}-a_{1}^{\dagger}a_{0}\right)\,,\\
\hat{L}_{-}&=-i\sum_{n=1}^{\infty}C_{n}\left(a_{n}^{\dagger}a_{n+1}+b_{n}^{\dagger}b_{n+1}\right)+iC_{0}\left(b_{1}a_{0}-a_{0}^{\dagger}a_{1}\right)\,.
\end{align}
\end{subequations}
The commutator between these charges is calculated using the anticommutation relations in Eq.~\eqref{CACRs} and the fact that $C_{n}^{2}-C_{n-1}^2=2n$. This is found to be
\begin{align}
\left[\hat{L}_{+},\hat{L}_{-}\right]=2\sum_{n=1}^{\infty}n\left(a_{n}^{\dagger}a_{n}+b_{n}^{\dagger}b_{n}\right)+C_{0}^2\mathbb{I}\,.
\end{align}
Comparing this with Eq.~\eqref{Comm-LadderOps}, we identify the right--hand side with $2\hat{L}_{0}$, where
\begin{align}
\hat{L}_{0}=\sum_{n=1}^{\infty}n\left(a_{n}^{\dagger}a_{n}+b_{n}^{\dagger}b_{n}\right)+\frac{1}{2}\left(\frac{1}{4}-M^2\right)\mathbb{I}\,,
\end{align}
is the time--translation charge operator. The canonical vacuum state $\ket{0}$ satisfies
\begin{align}
\hat{L}_{0}\ket{0}=\frac{1}{2}\left(\frac{1}{4}-M^2\right)\ket{0}\,,\hspace{.5cm}\hat{L}_{-}\ket{0}=0\,.
\end{align}
However, in contrast to the previous case, the vacuum sector with energy $(1/4-M^2)/2$ has a double degeneracy: The state $a_{0}^{\dagger}\ket{0}$ also satisfies $\hat{L}_{0}a_{0}^{\dagger}\ket{0}=\left((1/4-M^2)/2\right)a_{0}^{\dagger}\ket{0}$ and $\hat{L}_{-}a_{0}^{\dagger}\ket{0}=0$. This was indeed expected from the fact that there is a zero--frequency mode $\Psi_{0}^{I\!I\!I}$ in the solution space. We therefore have a two--parameter family of (normalized) vacuum states, given by
\begin{align}
\ket{0;\alpha}:=\alpha\ket{0}+\left(1-|\alpha|^2\right)^{\frac{1}{2}}a_{0}^{\dagger}\ket{0}\,,\hspace{.5cm}\alpha\in\mathbb{C}\,,
\end{align}
which implies that for every $\alpha\in\mathbb{C}$, the vacuum sector for these theories generates the lowest weight module isomorphic to the discrete series representation $\mathscr{D}_{(1/4-M^2)/2}^{+}$.

\section{Conclusion}\label{Concl}

In this paper we studied the solutions of the Dirac equation with mass $M$ in two-dimensional universal cover of anti-de~Sitter space, $\mathrm{AdS}_2$. We first determined all possible boundary conditions at the spatial boundary based on the requirement that the operator $\mathbb{D}$, defined by Eq.~\eqref{OpD},  should be extended to a self--adjoint operator following the general theory of Weyl~\cite{weyl} and von~Neumann.~\cite{neu} Since a solution of the Dirac equation with mass $-M$ can be obtained from a solution $\Psi$ with mass $M$ by $\Psi\mapsto-2i\Sigma^{01}\Psi$, we can restrict the mass values to $M\geq 0$. 

For $M\geq 1/2$ the self--adjoint extension of the operator $\mathbb{D}$ is unique and determined to correspond to the Dirichlet type I boundary condition for the components $\Phi^{(1)}$, $\Phi^{(2)}$ of the associated spatial spinor $\Phi$ at the endpoints $\rho=\pm\pi/2$ of the form given by Eq.~\eqref{aux2}. For $0 \leq M < 1/2$ the self-adjoint extensions of $\mathbb{D}$ are labeled by a $2\times 2$ unitary matrix $U$, which parametrizes the boundary conditions as in Eq.~\eqref{SABC1}.

Next, we determined the self--adjoint boundary conditions which are invariant under the action of the group $\widetilde{\mathrm{SL}}(2,\mathbb{R})$, which is the symmetry group of $\mathrm{AdS}_2$. For $0 < M < 1/2$, we found that the only unitary matrices $U$ parametrizing the boundary conditions which result in invariant mode solutions are given by $U=\mathrm{diag}(\mp 1,\pm 1)$ and $U=\pm\mathbb{I}$. These matrices correspond to the Dirichlet boundary conditions of type I, II, III and IV, respectively, defined in Eqs.~\eqref{Bc1}--\eqref{Bc4}. For the massless case $M=0$, we found that any diagonal unitary matrix $U$ gives a set of boundary conditions that result in invariant mode solutions. We also noted that the Dirichlet type I--IV boundary conditions for all $0\leq M<1/2$ and the Dirichlet type I boundary condition for all $M\geq 0$ are invariant under charge conjugation. 

The set of solutions to the Dirac equation satisfying an invariant boundary condition forms a unitary representation of the group $\widetilde{\mathrm{SL}}(2,\mathbb{R})$, but this representation may not split into invariant positive-- and negative--frequency subspaces, which is necessary for the vacuum state of the quantized theory to be isometry--invariant as shown in Appendix~\ref{App2}. We found that the positive--frequency solutions span invariant subspaces for the Dirichlet type I boundary condition for all $M\geq 0$, while the Dirichlet type II condition lead to invariant positive--frequency subspaces only for $0\leq M<1/2$. The mode functions resulting from these boundary conditions were identified with the sum of discrete series representations, $\mathscr{D}_{1/2+M}^{+}\oplus\mathscr{D}_{1/2+M}^{-}$ for the Dirichlet type I modes and  $\mathscr{D}_{1/2-M}^{+}\oplus\mathscr{D}_{1/2-M}^{-}$ for the Dirichlet type II modes. Both Dirichlet types III and IV mode functions are identified with the complementary series representations $\mathscr{C}_{1/2+M}^{0}$ which are already irreducible and do not split into invariant positive-- and negative--frequency subspaces.  For the massless case we found that the only diagonal unitary matrices corresponding to boundary conditions that result in invariant positive-- and negative--frequency subspaces are of the form $U=\mathrm{diag}(e^{2i\beta_{+}},-e^{-2i\beta_{+}})$, with $\beta_{+}\in[0,\pi)$. The particular cases for  $\beta_{+}=\pi/2$ and $\beta_{+}=0$ correspond to the massless Dirichlet conditions of type I and II, respectively. We noted that all the other massless mode solutions that form invariant positive--frequency subspaces are actually related to the Dirichlet type I and II mode solutions by a chiral transformation realized as the action of the rotation by the angle $\pi/2-\beta_{+}$ on the spatial components of these modes. Because the massless Dirac equation is invariant under chiral transformations, the solutions parametrized by $\beta_{+}$ are taken to be equivalent, and thus, can be identified with the Dirichlet type I (or type II) mode solutions. These mode solutions, up to a chiral transformation, are identified with the sum of mock--discrete series representations $\mathscr{D}_{1/2}^{+}\oplus\mathscr{D}_{1/2}^{-}$. For all other diagonal matrices, the associated self--adjoint boundary condition results in mode functions forming the principal series representation $\mathscr{P}_{0}^{\mu}$, where $\mu$ depends on the parameters $\beta_{\pm}$ via Eq.~\eqref{MuParam}. It is worth pointing out that all mode solutions that we obtained from the $\SLC$--invariant self--adjoint extensions of the operator $\mathbb{D}$ \emph{i.e.}, the Dirichlet type I modes for all $M\geq 0$ and Dirichlet types I--IV for $0\leq M<1/2$, are precisely the modes that Sakai and Tanii found through the requirement that the energy flux at $\rho=\pm\pi/2$ vanishes separately. It will be interesting to investigate deeper connections, if any, between these two requirements.

Finally, we examined the cases for which the self--adjoint boundary conditions are $\SLC$--invariant but the solution spaces do not split into invariant positive-- and negative--frequency subspaces, \emph{i.e.}, the massless solution spaces satisfying the boundary conditions with $\beta_{+}+\beta_{-}\neq\pi/2,3\pi/2$, and the massive solution spaces satisfying the Dirichlet type III and IV boundary conditions. Due to the lack of an invariant positive--frequency subspace, the vacuum state associated to these theories is not invariant, but instead belongs to a UIR of $\SLC$. For the massless theories, we found that the vacuum state belongs to the discrete series representation $\mathscr{D}_{\lambda}^{+}$, with $\lambda=(\mu-1/2)^{2}/2$, $\mu$ given by Eq.~\eqref{MuParam}. The massless theory with $\mu=0$, and the massive theories corresponding to the Dirichlet type III and type IV mode solutions resulted in a doubly degenerate vacuum sector. The UIR to which the vacuum state $\ket{0;\alpha}$, for $\alpha\in\mathbb{C}$ belongs is isomorphic to the discrete series $\mathscr{D}_{(1/4-M^2)/2}^{+}$. 

\acknowledgments

The author wishes to express profound gratitude to his thesis supervisor Prof.~Atsushi Higuchi, under whose direction this work was done, for his useful comments on earlier versions of this paper. The author also thanks Vasileios Letsios for useful discussions. This work was supported by an Overseas Research Studentship from the University of York.

\appendix

\section{Free Dirac field in static spacetime}\label{App2}

In this section we review a non--interacting Dirac field in a static spacetime with a stationary vacuum state.~\cite{waldQFTCSbook,weiminjin} We explain that if the space of solutions of the Dirac equation split into positive-- and negative--frequency subspaces invariant under the isometry group of the theory, then the vacuum state resulting from the Fock space construction after canonical quantization is invariant.

Consider a $D$--dimensional static spacetime $\mathcal{M}$ with local coordinates $(x^{\mu})=(t,x)$, $\mu=0,1,\dots,D-1$, and metric given by
\begin{align}\label{StaticMetric}
\D s^2=-N(x)^2\D t^2+g_{ij}(x)\D x^{i}\D x^{j}\,,
\end{align}
with $i,j=1,\dots,D-1$. The non-zero components of the Levi--Civita connection are found to be
\begin{subequations}\label{StaticConnComp}
\begin{align}
\Gamma^{0}_{\,\,i0}&=\frac{1}{N}\partial_{i}N\,,\\
\Gamma^{i}_{\,\,00}&=g^{ij}N\partial_{j}N\,,
\end{align}
\end{subequations}
and the $\Gamma^{i}_{\,\,jk}$ are the associated connection components of the $(D-1)$--dimensional Riemannian metric $g_{ij}$. We can construct a time--independent orthonormal frame $\{e_{a}\}_{a=0}^{D-1}$ with internal flat Lorentzian metric $(\eta_{ab})=\mathrm{diag}(-1,1,\dots,1)$ by defining the component functions $e_{a}^{\,\,\,\mu}$ through Eq.~\eqref{zweibein} and letting $e_{0}^{\,\,\,0}=1/N(x)$, together with $e_{0}^{\,\,\,i}=0=e_{i}^{\,\,\,0}$ and with $e_{i}^{\,\,\,j}$ functions of $x$. With respect to this frame, the connection $1$--form $\omega^{a}_{\,\,\,b}$ is once again defined by Eq.~\eqref{spinconnection}, and it is found that the relevant non-zero components $\omega_{ab\,\mu}$ are given by
\begin{subequations}\label{appConnection}
\begin{align}
\omega_{0i\,0}&=-e_{i}^{\,\,\,j}\partial_{j}N(x)\,,\\
\omega_{ij\,k}&=\left(\tilde{\nabla}_{k}e_{j}^{\,\,\,l}\right)e_{il}\,,
\end{align}
\end{subequations}
where $\tilde{\nabla}$ here stands for the covariant derivative with respect to the connection $(\Gamma^{i}_{\,jk})$. We use any representation for the $D$ gamma matrices $\gamma^{a}$ for which $(\gamma^{0})^{\dagger}=-\gamma^{0}$ and $(\gamma^{i})^{\dagger}=\gamma^{i}$. We can now find the spinor covariant derivative as given by Eq.~\eqref{covariantder}, which reads
\begin{subequations}\label{appCovariant}
\begin{align}
\nabla_{0}&=\partial_{0}-e_{i}^{\,\,\,j}\partial_{j}N(x)\Sigma^{0i}\,,\\
\nabla_{i}&=\partial_{i}+\frac{1}{2}\left(\tilde{\nabla}_{i}e_{k}^{\,\,\,l}\right)e_{jl}\Sigma^{jk}\,,
\end{align}
\end{subequations}
where as before, $\Sigma^{ab}=[\gamma^{a},\gamma^{b}]/4$. The Lagrangian for a free Dirac field $\psi\in C^{\infty}(\mathcal{M},\mathbb{C}^{\tilde{D}})$ where $\tilde{D}=2^{\lfloor D/2\rfloor}$, with mass $M$ and Dirac adjoint $\psi^{*}=\psi^{\dagger}\gamma^{0}$, is thus given by
\begin{subequations}\label{DiracLagr}
\begin{align}
L&=\int_{\Sigma}\mathscr{L}\D x\,,\\
\mathscr{L}&=N\sqrt{g}\,\psi^{*}\left(\tilde{\gamma}^{\mu}\nabla_{\mu}-M\right)\psi\,,
\end{align}
\end{subequations}
where $g:=\mathrm{det}(g_{ij})$ and where we have defined $\tilde{\gamma}^{\mu}=e_{a}^{\,\,\,\mu}\gamma^{a}$. Here, $\Sigma$ is a spacelike hypersurface of constant $t$. The conjugate momentum density is found to be 
\begin{align}\label{MomentumDen}
\pi(t,x)&=\frac{\delta\mathscr{L}}{\delta(\nabla_{0}\psi(t,x))}\nonumber\\
&=i\sqrt{g(x)}\,\psi^{\dagger}(t,x)\,,
\end{align}
where we have used the left functional derivative.
The $U(1)$ transformation $\psi\mapsto e^{i\theta}\psi$, $\theta\in\mathbb{R}$ leaving the Lagrangian density~\eqref{DiracLagr} invariant has the associated Noether current $\psi^{*}\tilde{\gamma}^{\mu}\psi$, with conserved charge
\begin{align}\label{NoetherCharge}
\int_{\Sigma}\overline{\psi}\gamma^{0}\psi\,e^{\,\,\,0}_{0}(x)N(x)\sqrt{g(x)}\,\D x=\int_{\Sigma}\psi^{\dagger}\psi\sqrt{g(x)}\,\D x\,.
\end{align}
Now, writing the components of the spinor $\psi$ with respect to the frame $\{e_{a}\}$ as  $\psi_{a}$, the equal time canonical anticommutation relations then read
\begin{subequations}\label{CARs}
\begin{align}
&\left\{\psi_{a}(t,x),\psi_{b}^{\dagger}(t,y)\right\}=\frac{1}{\sqrt{g(x)}}\delta_{ab}\delta(x,y)\,,\\
&\left\{\psi_{a}(t,x),\psi_{b}(t,y)\right\}=0=\left\{\psi_{a}^{\dagger}(t,x),\psi_{b}^{\dagger}(t,y)\right\}\,,
\end{align}
\end{subequations}
where
\begin{align}
\int_{\Sigma}\delta(x,y)f(y)\D y=f(x)\,,
\end{align}
for any smooth compactly supported function $f$ on $\Sigma$.

The Euler--Lagrange equations resulting from the Lagrangian in Eq.~\eqref{DiracLagr} reduce to the Dirac equation of the form given in Eq.~\eqref{DiracEq1}. It can be written using Eqs.~\eqref{appConnection} and~\eqref{appCovariant} in this spacetime as
\begin{align}
\partial_{t}\psi(t,x)=\left[\tilde{\gamma}^{0}\tilde{\gamma}^{i}\left(N(x)^2\nabla_{i}+\frac{1}{2}N(x)\partial_{i}N(x)\right)-MN(x)^2\tilde{\gamma}^{0}\right]\psi(t,x)\,.
\end{align}
We now assume that the spinor $\psi$ is  of the form
\begin{align}
\psi(t,x)=\Phi_{\sigma}(x)e^{-i\omega_{\sigma}t}\,,
\end{align}
where the spatial component spinor $\Phi_{\sigma}$ is a solution to the equation
\begin{align}\label{appDirEq}
\mathbb{D}_D\Phi_{\sigma}(x)=\omega_{\sigma}\Phi_{\sigma}(x)\,,
\end{align}
where we have defined the operator
\begin{align}
\mathbb{D}_{D}=iN(x)^2\tilde{\gamma}^{0}\left[\tilde{\gamma}^{i}\left(\nabla_{i}+\frac{1}{2}N(x)^{-1}\partial_{i}N(x)\right)-M\right]\,.
\end{align}

We note that the operator $\mathbb{D}_{D}$ satisfies
\begin{align}\label{symmetric}
\inpr{\Phi_1}{\mathbb{D}_{D}\Phi_2}=\inpr{\mathbb{D}_{D}\Phi_1}{\Phi_2}\,,
\end{align}
where the inner product has been defined through the Noether charge in Eq.~\eqref{NoetherCharge} as
\begin{align}\label{appInpr}
\inpr{\Phi_1}{\Phi_2}:=\int_{\Sigma}\Phi_{1}(x)^{\dagger}\Phi_{2}(x)\sqrt{g(x)}\D x\,,
\end{align}
whenever the boundary term vanishes.

Suppose now, that the operator $\mathbb{D}_{D}$ defined on an appropriate domain is self--adjoint with respect to the inner product~\eqref{appInpr}, \emph{i.e.}, that Eq.~\eqref{symmetric} is satisfied and the domain of the adjoint operator is the same as the domain of $\mathbb{D}_{D}$. Suppose further that the spectrum of $\mathbb{D}_{D}$ is discrete and $\omega_{\sigma}^2>0$. This implies that if $\Phi_{\sigma}$ is a solution of Eq.~\eqref{appDirEq} with $\omega_{\sigma}>0$, then a solution of this equation with $\omega_{\sigma}\to -\omega_{\sigma}$ is given by $\Phi_{\sigma}^{c}:=C\left(\Phi_{\sigma}^{*}\right)^{T}$, with the charge conjugation matrix $C$ defined through~\cite{pal} $C^{-1}\gamma^{a}C=-(\gamma^{a})^{T}$, $C^{-1}=C^{\dagger}$ and, hence, the solution space splits into positive-- and negative--frequency subspaces. Then, the eigenspinors of $\mathbb{D}_D$ form a complete set, and thus, the quantum field can be expanded as
\begin{align}\label{Qfield}
\psi(t,x)=\sum_{\sigma}\left[a_{\sigma}\Phi_{\sigma}(x)e^{-i\omega_{\sigma}t}+b_{\sigma}^{\dagger}\left(\Phi_{\sigma}(x)\right)^{c}e^{i\omega_{\sigma}t}\right]\,.
\end{align}

Eq.~\eqref{symmetric} implies that if $\omega_{\sigma}\neq\omega_{\sigma'}$, then the spinors $\Phi_{\sigma}$ and $\Phi_{\sigma'}$ are orthogonal. This allows us to normalize the mode solutions by imposing
\begin{align}
\inpr{\Phi_\sigma}{\Phi_{\sigma'}}=\delta_{\sigma\sigma'}\,,
\end{align}
which, together with the fact that the set of solutions is complete, in turn implies that
\begin{align}\label{deltacomplete}
\sum_{\sigma}\Phi_{\sigma}(x')^{\dagger}\Phi_{\sigma}(x)=\frac{1}{\sqrt{g(x)}}\delta(x,x')\,.
\end{align}
This relation is used to show that the anticommutation relations between the fields $\psi$ and $\psi^{\dagger}$ in Eq.~\eqref{CARs} are equivalent to the anticommutation relations between the creation and annihilation operators given by
\begin{align}\label{CARs2}
\left\{a_{\sigma},a^{\dagger}_{\sigma'}\right\}=\delta_{\sigma\sigma'}=\left\{b_{\sigma},b^{\dagger}_{\sigma'}\right\}\,,
\end{align}
with all the other anticommutators vanishing. Then, the Fock space can be constructed by defining the vacuum state $\ket{0}$ requiring that
\begin{align}\label{vacuumstate}
a_{\sigma}\ket{0}=0=b_{\sigma}\ket{0}\,,
\end{align}
is satisfied for all $\sigma$.

Now, let $\xi$ be a Killing vector of the spacetime, and let $\mathcal{L}_{\xi}$ be the spinorial Lie derivative in  the direction of $\xi$, defined by Eq.~\eqref{LieDer}. Then,
\begin{align}
\mathcal{L}_{\xi}\left[\Phi_{\sigma}(x)e^{-i\omega_{\sigma}t}\right]=\sum_{\sigma'}\left(\Lambda_{\sigma\sigma'}\Phi_{\sigma'}(x)e^{-i\omega_{\sigma'}t}+\tilde{\Lambda}_{\sigma\sigma'}\left(\Phi_{\sigma'}(x)\right)^{c}e^{i\omega_{\sigma'}t}\right)\,,
\end{align}
so that after substituting the above into Eq.~\eqref{Qfield}, we have that
\begin{align}
\mathcal{L}_{\xi}\psi(t,x)=&\sum_{\sigma'}\sum_{\sigma}\left[\left(a_{\sigma}\Lambda_{\sigma\sigma'}+b_{\sigma}^{\dagger}\overline{\tilde{\Lambda}_{\sigma\sigma'}}\right)\Phi_{\sigma'}(x)e^{-i\omega_{\sigma'}t}\right.\nonumber\\
&+\left.\left(a_{\sigma}\tilde{\Lambda}_{\sigma\sigma'}+b_{\sigma}^{\dagger}\overline{\Lambda_{\sigma\sigma'}}\right)\Phi_{\sigma'}(x)^{c}e^{i\omega_{\sigma'}t}\right]\,,
\end{align}
where we have used the fact that $\left(\mathcal{L}_{\xi}\psi\right)^{c}=\mathcal{L}_{\xi}\psi^{c}$. Thus, we see that the Lie derivative on $\psi$ induces the transformation on the operators
\begin{subequations}
\begin{align}
a_{\sigma}&\mapsto\sum_{\sigma'}\left(a_{\sigma'}\Lambda_{\sigma'\sigma}+b_{\sigma'}^{\dagger}\overline{\tilde{\Lambda}_{\sigma'\sigma}}\right)\,,\\
b_{\sigma}^{\dagger}&\mapsto\sum_{\sigma'}\left(a_{\sigma'}\tilde{\Lambda}_{\sigma'\sigma}+b_{\sigma'}^{\dagger}\overline{\Lambda_{\sigma'\sigma}}\right)\,.
\end{align}
\end{subequations}
Hence, for the vacuum state $\ket{0}$ defined by Eq.~\eqref{vacuumstate} to be invariant under the spacetime symmetry transformation
corresponding to the Killing vector $\xi$, we need to have $\tilde{\Lambda}_{\sigma\sigma'}=0$. That is,
\begin{align}
\mathcal{L}_{\xi}\left[\Phi_{\sigma}(x)e^{-i\omega_{\sigma}t}\right]=\sum_{\sigma'}\Lambda_{\sigma\sigma'}\Phi_{\sigma'}(x)e^{-i\omega_{\sigma'}t}\,.
\end{align}
In other words, for the vacuum state $\ket{0}$ to be invariant under this symmetry transformation, the positive--frequency solutions must transform among themselves without any component of negative--frequency solutions. Finally, we note that $\ket{0}$ is stationary, \emph{i.e.} invariant under time translation symmetry $T$ with $\mathcal{L}_{T}\psi=\partial_{t}\psi$.

\section{Asymptotic behavior of Ferrers functions}\label{App1}

In this section we show that the asymptotic behavior at the spatial boundary of the component functions in Eq.~\eqref{SolCompklog} corresponding to Dirac spinors of mass $M=k+1/2$ ($k\in\mathbb{N}\cup\{0\}$), is given by Eqs.~\eqref{SolCompkAsymp-plus} and~\eqref{SolCompkAsymp-minus} if $k>0$, and we show Eqs.~\eqref{Squarek0Asymp} for the case $k=0$.

We begin by analyzing the case $k>0$. The behavior of the Ferrers functions $\mathsf{P}_{\omega}^{-k}(x)$ and $\mathsf{Q}_{\omega}^{-k}$, for $\omega\in\mathbb{C}$ with $\omega\neq 0$, at the singular point $x=1$ is given by~\cite{nist}
\begin{subequations}\label{Ferrers-asym1}
\begin{align}
\mathsf{P}_{\omega}^{-k}(x)&\approx\frac{1}{\Gamma(1+k)}\left(\frac{1-x}{2}\right)^{\frac{k}{2}}\,,\label{FerrersP-asym1}\\
\mathsf{Q}_{\omega}^{-k}(x)&\approx \frac{\Gamma(k)\Gamma(\omega-k+1)}{2\Gamma(\omega+k+1)}\left(\frac{2}{1-x}\right)^{\frac{k}{2}}\,,\label{FerrersQ-asym1}
\end{align}
\end{subequations}
where $f(x)\approx g(x)$ if and only if $f(x)/g(x)\to 1$ as $x\to c$. For sufficiently small $\epsilon>0$, we have that if $\rho=\pi/2-\epsilon$, then $\sin\rho=\cos\epsilon$, thus, Eqs.~\eqref{Ferrers-asym1} with $x=\cos\epsilon$ imply that
\begin{subequations}
\begin{align}
\mathsf{P}_{\omega}^{-k}(\cos\epsilon)\pm\mathsf{P}_{\omega-1}^{-k}(\cos\epsilon)&\approx  0\,,\label{FerrersPdiff-asym}\\
\mathsf{Q}_{\omega}^{-k}(\cos\epsilon)+\mathsf{Q}_{\omega-1}^{-k}(\cos\epsilon)&\approx \frac{2^{k}\omega\Gamma(\omega-k)\Gamma(k)}{\Gamma(\omega+k+1)} \epsilon^{-k}\,,\label{FerrersQdiff-asymPlus}\\
\mathsf{Q}_{\omega}^{-k}(\cos\epsilon)-\mathsf{Q}_{\omega-1}^{-k}(\cos\epsilon)&\approx \frac{2^{k}k\Gamma(\omega-k)\Gamma(k)}{\Gamma(\omega+k+1)} \epsilon^{-k}\,,\label{FerrersQdiff-asymMin}
\end{align}
\end{subequations}
where we have used the estimate $\cos\epsilon\sim 1-\epsilon^2/2$ so that $(1-\cos\epsilon)^{-k/2}\sim \epsilon^{-k}$. Noting that for the same value of $\rho$, we have
\begin{align}\label{estimate1}
\left(\frac{1-\sin\rho}{1+\sin\rho}\right)^{\frac{1}{4}}\sim \epsilon^{\frac{1}{2}}\,.
\end{align}
Then it follows that the component functions in Eq.~\eqref{SolCompklog} evaluated at $\rho=\pi/2-\epsilon$ behave as
\begin{subequations}
\begin{align}
\Phi_{\omega}^{(1)}\left(\frac{\pi}{2}-\epsilon\right)&\sim C_2 \frac{2^{k}\omega\Gamma(\omega-k)\Gamma(k)}{\Gamma(\omega+k+1)} \epsilon^{-k+\frac{1}{2}}\,,\\
\Phi_{\omega}^{(2)}\left(\frac{\pi}{2}-\epsilon\right)&\sim C_2 \frac{2^{k}k\Gamma(\omega-k)\Gamma(k)}{\Gamma(\omega+k+1)} \epsilon^{-k-\frac{1}{2}}\,,
\end{align}
\end{subequations}
which are precisely the expressions in Eq.~\eqref{SolCompkAsymp-plus}.

To analyze the behavior of the component functions $\Phi_{\omega}^{(1)}$ and $\Phi_{\omega}^{(2)}$ at the other endpoint $\rho=-\pi/2$, we use the connection formulas for the Ferrers functions,~\cite{nist} namely
\begin{align}\label{connectionfor}
\mathsf{P}_{\omega}^{-k}(-x)&=\cos\pi(\omega-k)\mathsf{P}_{\omega}^{-k}(x)-\frac{2}{\pi}\sin\pi(\omega-k)\mathsf{Q}_{\omega}^{-k}(x)\,,\nonumber\\
\mathsf{Q}_{\omega}^{-k}(-x)&=-\cos\pi(\omega-k)\mathsf{Q}_{\omega}^{-k}(x)-\frac{\pi}{2}\sin\pi(\omega-k)\mathsf{P}_{\omega}^{-k}(x)\,,
\end{align}
from which it follows that
\begin{subequations}\label{connect-plusmin}
\begin{align}
\mathsf{P}_{\omega}^{-k}(-x)\pm\mathsf{P}_{\omega-1}^{-k}(-x)=&\cos\pi(\omega-k)\left(\mathsf{P}_{\omega}^{-k}(x)\mp\mathsf{P}_{\omega-1}^{-k}(x)\right)\nonumber\\
&-\frac{2}{\pi}\sin\pi(\omega-k)\left(\mathsf{Q}_{\omega}^{-k}(x)\mp\mathsf{Q}_{\omega-1}^{-k}(x)\right)\,,\\
\mathsf{Q}_{\omega}^{-k}(-x)\pm\mathsf{Q}_{\omega-1}^{-k}(-x)=&-\cos\pi(\omega-k)\left(\mathsf{Q}_{\omega}^{-k}(x)\mp\mathsf{Q}_{\omega-1}^{-k}(x)\right)\nonumber\\
&-\frac{\pi}{2}\sin\pi(\omega-k)\left(\mathsf{P}_{\omega}^{-k}(x)\mp\mathsf{P}_{\omega-1}^{-k}(x)\right)\,.
\end{align}
\end{subequations}
Taking the plus sign above and setting $\rho=\epsilon-\pi/2$, so that $-x=\sin\rho=-\cos\epsilon$, and substituting Eqs.~\eqref{FerrersPdiff-asym} and~\eqref{FerrersQdiff-asymMin} we obtain
\begin{subequations}\label{FerrersAux}
\begin{align}
\mathsf{P}_{\omega}^{-k}(-\cos\epsilon)+\mathsf{P}_{\omega-1}^{-k}(-\cos\epsilon)
&\approx \sin\pi(\omega-k)\frac{2^{k+1}k\Gamma(\omega-k)\Gamma(k)}{\pi\Gamma(\omega+k+1)} \epsilon^{-k}\,,\\
\mathsf{Q}_{\omega}^{-k}(-\cos\epsilon)+\mathsf{Q}_{\omega-1}^{-k}(-\cos\epsilon)&\approx\cos\pi(\omega-k)\frac{2^{k}k\Gamma(\omega-k)\Gamma(k)}{\Gamma(\omega+k+1)} \epsilon^{-k}\,,
\end{align}
\end{subequations}
therefore, using the analogous estimate appearing in Eq.~\eqref{estimate1} this time with $\rho\to -\rho$, it follows that the behavior of the component function $\Phi_{\omega}^{(1)}$  is found to be
\begin{align}\label{ApResult1}
\Phi_{\omega}^{(1)}\left(\epsilon-\frac{\pi}{2}\right)\approx & \frac{2^{k}k\Gamma(\omega-k)\Gamma(k)}{\Gamma(\omega+k+1)}\left[C_1\frac{2}{\pi}\sin\pi(\omega-k)+C_2\cos\pi(\omega-k)\right]\epsilon^{-k-\frac{1}{2}}\,.
\end{align}
Similarly, taking the minus sign in Eq.~\eqref{FerrersAux} and $\rho=\epsilon-\pi/2$ we obtain
\begin{subequations}\label{FerrersAux2}
\begin{align}
\mathsf{P}_{\omega}^{-k}(-\cos\epsilon)-\mathsf{P}_{\omega-1}^{-k}(-\cos\epsilon)
&\approx \sin\pi(\omega-k)\frac{2^{k+1}\omega\Gamma(\omega-k)\Gamma(k)}{\pi\Gamma(\omega+k+1)} \epsilon^{-k}\,,\\
\mathsf{Q}_{\omega}^{-k}(-\cos\epsilon)-\mathsf{Q}_{\omega-1}^{-k}(-\cos\epsilon)&\approx\cos\pi(\omega-k)\frac{2^{k}\omega\Gamma(\omega-k)\Gamma(k)}{\Gamma(\omega+k+1)} \epsilon^{-k}\,,
\end{align}
\end{subequations}
and thus, the second component $\Phi_{\omega}^{(2)}$ satisfies 
\begin{align}\label{ApResult2}
\Phi_{\omega}^{(2)}\left(\epsilon-\frac{\pi}{2}\right)\approx & \frac{2^{k}\omega\Gamma(\omega-k)\Gamma(k)}{\Gamma(\omega+k+1)} \left[\frac{2}{\pi}C_1\sin\pi(\omega-k)+C_2\cos\pi(\omega-k)\right]\epsilon^{-k+\frac{1}{2}}\,.
\end{align}
Equations~\eqref{ApResult1} and~\eqref{ApResult2} are Eq.~\eqref{SolCompkAsymp-minus}.

For the case when $k=0$, the Ferrers functions in the component functions of Eq.~\eqref{SolCompklog} reduce to Legendre functions of the form $\mathsf{P}_{\omega}(x)$ and $\mathsf{Q}_{\omega}(x)$. At $x=1$ we have $P_{\omega}(1)=1$ for the functions of the first kind, and for the functions of the second kind we have, as $x\to 1$, 
\begin{align}\label{LogQ-Expan}
\mathsf{Q}_{\omega}(x)\approx \mathsf{P}_{\omega}(x)\left[\frac{1}{2}\ln\left(\frac{1+x}{1-x}\right)-\gamma-\psi(\omega+1)\right]\,,
\end{align}
where $\psi$ denotes the digamma function, related to the gamma function by $\psi(x)=\Gamma'(x)/\Gamma(x)$, and $\gamma$ the Euler constant.~\cite{nist} Letting  $\rho=\pi/2-\epsilon$, for sufficiently small values of $\epsilon>0$ we use Eq.~\eqref{LogQ-Expan} with $x=\cos\epsilon$ to obtain
\begin{subequations}\label{LogQ-diff}
\begin{align}
\mathsf{Q}_{\omega}(\cos\epsilon)+\mathsf{Q}_{\omega-1}(\cos\epsilon)&\approx \ln 2-2\ln\epsilon-2\gamma-2\psi(\omega)-\frac{1}{\omega}\,,\\
\mathsf{Q}_{\omega}(\cos\epsilon)-\mathsf{Q}_{\omega-1}(\cos\epsilon)&\approx -\frac{1}{\omega}\,,
\end{align}
\end{subequations}
where we have used the estimate $1-\cos\epsilon\sim\epsilon^2/2$ and the identity $\psi(\omega+1)=\psi(\omega)+1/\omega$. Using Eq.~\eqref{estimate1} we take into account the prefactors of the component functions in Eq.~\eqref{SolCompklog}, and thus we have 
\begin{subequations}\label{ApResult3}
\begin{align}
\Phi_{\omega}^{(1)}\left(\frac{\pi}{2}-\epsilon\right)&\approx -2\,C_2\, \epsilon^{\frac{1}{2}}\ln\epsilon\,,\\
\Phi_{\omega}^{(2)}\left(\frac{\pi}{2}-\epsilon\right)&\approx \frac{1}{\omega}C_2 \epsilon^{-\frac{1}{2}}\,,
\end{align}
\end{subequations}
hence, considering that $\epsilon^{1/2}\ln\epsilon\to 0$ as $\epsilon\to 0$, it is clear that Eq.~\eqref{Squarek0Asymp1} immediately follows.

To evaluate at the other endpoint of the spatial boundary, we use the connection formulas for the Legendre functions which are obtained by setting $k=0$ in Eq.~\eqref{connectionfor}. It then follows that the identities in Eq.~\eqref{connect-plusmin} with $k=0$ are valid for the Legendre functions. By setting $-x=\sin\rho$ with $\rho=\epsilon-\pi/2$, and using Eq.~\eqref{LogQ-diff} we have that
\begin{subequations}
\begin{align}
\mathsf{P}_{\omega}(-\cos\epsilon)+\mathsf{P}_{\omega-1}(-\cos\epsilon)&\approx \frac{2}{\pi\omega}\sin\pi\omega\,,\\
\mathsf{Q}_{\omega}(-\cos\epsilon)+\mathsf{Q}_{\omega-1}(-\cos\epsilon)&\approx \frac{1}{\omega}\cos\pi\omega\,,
\end{align}
\end{subequations}
for the plus sign, and
\begin{subequations}
\begin{align}
\mathsf{P}_{\omega}(-\cos\epsilon)-\mathsf{P}_{\omega-1}(-\cos\epsilon)\approx &-\frac{2}{\pi}\sin\pi\omega\left(2\ln \epsilon-\ln 2-2\gamma-2\psi(\omega)-\frac{1}{\omega}\right)\nonumber\\
&+2\cos\pi\omega\,,\\
\mathsf{Q}_{\omega}(-\cos\epsilon)-\mathsf{Q}_{\omega-1}(-\cos\epsilon)\approx & -\cos\pi\omega\left(2\ln\epsilon-\ln 2-2\gamma-2\psi(\omega)-\frac{1}{\omega}\right)\nonumber\\
&-\frac{\pi}{2}\sin\pi\omega\,,
\end{align}
\end{subequations}
for the minus sign. Once again, using Eq.~\eqref{estimate1} the expressions above imply that
\begin{subequations}
\begin{align}
\Phi_{\omega}^{(1)}\left(\epsilon-\frac{\pi}{2}\right)&\approx \left[C_1\frac{2}{\pi\omega}\sin\pi\omega+C_2\frac{1}{\omega}\cos\pi\omega\right]\epsilon^{-\frac{1}{2}}\,,\\
\Phi_{\omega}^{(2)}\left(\epsilon-\frac{\pi}{2}\right)&\approx -\left[C_1\frac{2}{\pi}\sin\pi\omega+\cos\pi\omega\right]\epsilon^{\frac{1}{2}}\ln\,\epsilon\,.
\end{align}
\end{subequations}
Using these approximations and the fact that $\epsilon^{1/2}\ln\epsilon\to 0$ as $\epsilon\to 0$, we have the asymptotic expansion given in Eq. \eqref{Squarek0Asymp2} immediately.

\bibliography{references}

\end{document}